\documentclass[aps,prl,onecolumn,superscriptaddress,preprintnumbers,showpacs]{revtex4}
\usepackage{graphics}
\def\eq#1\en{\begin{equation} #1 \end{equation}}

\def\eqa#1\ena{\begin{eqnarray} #1 \end{eqnarray}}

\usepackage{graphicx,color}

\begin{document}

\title{Moments of inertia, nucleon axial-vector coupling, 
the {\bf 8}, {\bf 10}, $\overline{\bf 10}$, and ${\bf 27}_{3/2}$ mass spectra
and the higher SU(3)$_f$ representation mass splittings in the Skyrme model}
\author{G. Duplan\v{c}i\'{c}}
\affiliation{Theoretical Physics Division, Rudjer Bo\v skovi\' c Institute, 
Zagreb, Croatia}
\author{H. Pa\v{s}agi\'{c}}
\affiliation{Faculty of Transport and Traffic Engineering, University of Zagreb, 
P.O. Box 195, 10000 Zagreb, Croatia} 
\author{J.Trampeti\'{c}}
\affiliation{Theoretical Physics Division, Rudjer Bo\v skovi\' c Institute, 
Zagreb, Croatia}
\affiliation{Theory Division, CERN, CH-1211 Geneva 23, Switzerland}
\affiliation{Theoretische Physik, Universit\"{a}t M\"{u}nchen, Theresienstr. 37, 80333 M\"{u}nchen, Germany}

\date{\today}

\begin{abstract}
The broad importance of a recent experimental discovery of 
pentaquarks requires more theoretical insight into
the structure of higher representation multiplets.
The nucleon axial-vector coupling, moments of inertia, 
the {\bf 8}, {\bf 10}, $\overline{\bf 10}$, and ${\bf 27}_{3/2}$ absolute mass spectra
and the higher SU(3)$_f$ representation mass splittings for the multiplets 
${\bf 8}$, ${\bf 10}$, $\overline{\bf 10}$, ${\bf 27}$, ${\bf 35}$, $\overline{\bf 35}$, and $\bf 64$
are computed in the framework of the minimal ${\rm SU(3)_f}$ extended Skyrme model
by using only one free parameter, i.e., the Skyrme charge $e$.
The analysis presented in this paper represents simple and clear theoretical estimates,
obtained without using any experimental results for higher ($\overline{\bf 10}$,...) multiplets.
The obtained results are in good agreement with other chiral soliton model approaches
that more extensively use experimental results as inputs.
\end{abstract}


\maketitle

\section{Introduction}

The experimental discovery \cite{penta1}--\cite{penta4} 
of the exotic baryon (probably spin 1/2) with strangeness +1, $\Theta^+$, 
was recently supported by the first observation of $\Theta^+$ 
in hadron--hadron interactions \cite{penta5},
and by the NA49 Collaboration \cite{penta6} discovery of the exotic isospin 3/2 baryon 
with strangeness -2, $ \Xi^{--}_{3/2}$. This discovery initiated huge interest in 
the theoretical high energy physics community. 
Namely, the antidecuplet, and possibly the other multiplets 
of the higher SU(3)$_{flavor}$ (SU(3)$_f$) representations, in this way
moved from pure theory into the real world of particle physics. 
The first successful prediction of mass of one member of the $\overline{\bf 10}$ baryons, 
known as penta-quark or $\Theta^+$-baryon, in the framework of the 
Skyrme model was presented in Ref. \cite{BD}.
To explain all other possible properties
concerning higher SU(3)$_f$ representations, like mass spectrums, relevant mass differences, etc.,
many authors used different types of chiral soliton 
\cite{MP,HB,park,dia1,wei1,WK,K,pra3,Itz,WM,Ell,DT,HW,DPT}, QCD \cite{Zhu},
quark \cite{Kar,Glo,Hua,Ger}, diquark \cite{JW,JW1,JW2}, 
lattice QCD \cite{lattice,lattice1,lattice2}, $1/N_c$ expansion \cite{JM1}
and many other methods and models \cite{DP,Bor,Bij,DPP,JM2}. 

The Skyrme model \cite{sky} has been very successful in providing a description of the so-called
long-distance properties of strong interactions.
Its QCD origin, beauty and simplicity is also a good motivation for reexamining 
the non-perturbative quantities, such as mass spectrums, baryon static properties, etc.
The idea of Skyrme \cite{sky} that baryons are solitons of an SU(2)$\times$SU(2) chiral theory
(or solitons in the non-linear sigma model), together with the 
't Hooft--Witten conjecture \cite{tho,witt}, attracted a lot of attention 
 \cite{bal}--\cite{gua}, went beyond all original expectations and 
developed into a remarkable theory \cite{wei}. 
 Since QCD (unlike QED) does not contain a natural
expansion parameter, 't Hooft \cite{tho} investigated the possibility of using $1/N_c$ as the expansion 
parameter, just as $\alpha$ is used in QED.
Following 't Hooft's argument, Witten \cite{witt} found what is today
known as the 't Hooft--Witten conjecture: ``As $N_c \to \infty$ 
QCD may be approximated at low energies
by a weakly coupled field theory of mesons, with baryons identified as topological soliton solutions''.
It is known that a topological feature of such a model is crucial:
``The topological number is interpreted as the baryon number''  \cite{sky}. In the 
$N_c \to \infty$ limit, baryons appear to be some kind of solitons 
in the effective mesonic field
theory \cite{witt}. Note that the anomalous baryon number current obtained from 
the Wess--Zumino term \cite{wess}
using the method of Goldstone and Wilczek \cite{gol} is still present in the SU(2)$_f$ case.
It is also interesting that the precise notion of the functional integral 
for a sector of a given fermion
number makes possible an exact proof for direct connection between baryons of 
QCD and solitons of the non-linear 
sigma model \cite{sarat}.

One can simply say that in this type of models, baryons emerge as soliton configurations of 
pseudoscalar mesons. Extension of the model to the strange sector, in order 
to account for a large strange quark mass,
requires that appropriate chiral-symmetry breaking terms should be included.
Also the scale invariant Wess--Zumino term \cite{wess,witt} has to be 
included into the total action ${\cal L}$
to obtain a configuration with the necessary constraint on the hypercharge $Y=1$ \cite{gua,MP,pra3,wei1,WK}. 
The resulting effective Hamiltonian can 
be treated by starting from a flavor symmetric formulation in which existing 
kaon fields arise from rigid rotations
of the classical pion field. The associated collective coordinates, 
which parameterize these amplitude 
fluctuations of the soliton, are canonically quantized 
to generate states that possess the quantum numbers of 
physical strange baryons \cite{wei}. It turns out that the resulting 
collective Hamiltonian can be diagonalized exactly, 
even in the presence of flavor symmetry breaking \cite{yab,dia,toy}. 

Huge theoretical interest induced by the recent discovery of 
higher SU(3)$_f$ representation baryon states (penta-quarks)
is our main motivation to revisit the minimal SU(3)$_f$ extended Skyrme model,
which uses only one free parameter, the Skyrme charge $e$,
the only one flavor symmetry breaking (SB) term, proportional to $\lambda_8$ in the kinetic 
and the mass terms and the SU(2)$_f$ arctan ansatz embedded into the SU(3)$_f$ symmetry
as the simplest analitycal solution of the Euler--Lagrange equation. 

We applied recently that model to nonleptonic hyperon and 
$\Omega^-$decays \cite{dppt}--\cite{prat} and to exotic baryon mass
splittings and mass spectrum \cite{DT,DPT} producing reasonable agreement with experiments. 
In this paper we use the minimal SU(3)$_f$ extended Skyrme model and calculate 
the nucleon etc. static properties and the SU(3)$_f$ representations mass spectrums and relevant mass splittings,
as functions of the Skyrme charge $e$.

Our other motivations are as follows:
\begin{itemize}
\item Study of classical soliton mass ${\cal M}_{csol}$, in the framework of 
the minimal SU(3)$_f$ extended Skyrme model
with SU(3)$_f$ arctan ansatz, makes possible to find the new analytical expression for 
dimensionless size of the skyrmion, $x'_0$, as a function of the Skyrme charge $e$ 
and the SU(3)$_f$ symmetry breaking terms.
This new $x'_0$ quantity describes analytically the internal dynamic of SU(3)$_f$ 
symmetry breaking which takes place within skyrmion.
\item To find the range of values of $e$, with or without the SU(3)$_f$ symmetry breaking effects included,
which reasonably fits the experimental data
for the nucleon axial-vector coupling, moments of inertia,
the {\bf 8}, {\bf 10}, $\overline{\bf 10}$ and ${\bf 27}_{3/2}$ absolute mass spectrums
and the higher SU(3)$_f$ representation mass splittings $\Delta_{1,...,12}$, for the multiplets 
${\bf 8}$, ${\bf 10}$, $\overline{\bf 10}$, ${\bf 27}$, ${\bf 35}$, $\overline{\bf 35}$ and $\bf 64$.
From a quark model point of view the minimal SU(3)$_f$ multiplets 
{\bf 8} and {\bf 10} contain no additional 
$q\bar q$ pair. However, the families of penta-quarks ($\overline{\bf 10}$, {\bf 27}, {\bf 35}) 
and septu-quarks ({\bf 28}, $\overline{\bf 35}$, {\bf 64}) 
contain additional one and two $q\bar q$ pairs, respectively.
\item The advantage of the Skyrme model over the quark models, or vice versa, 
for correct description of higher
SU(3)$_f$ representation of baryons, i.e. description of penta-quark, etc. states,
would also become more transparent. In this context the evaluation of nucleon $g_A$ serves only as a 
consistency check of our approach as a whole.
\end{itemize}

The paper is organized as follows:
\begin{itemize}
\item First, we describe the basic features of the Skyrme 
model including the Hamiltonian, kinematics and the quantization procedure \cite{wit},
and introduce the SU(2)$_f$ arctan ansatz as the profile function
and recalculate nucleon static properties. 
\item Next is the construction of Noether currents and 
the introduction of an arctan ansatz as the profile function,
for the case of broken SU(3)$_f$ symmetry. The nucleon axial-vector coupling, moments of inertia, 
the {\bf 8}, {\bf 10}, $\overline{\bf 10}$ and ${\bf 27}_{3/2}$ absolute mass spectrums
and the higher SU(3)$_f$ representation mass splittings $\Delta_{1,...,12}$,
as functions of $e$ ($3\leq \,e\,\leq 5$) and the SU(3)$_f$ symmetry breaking
parameters ($m_\pi,\,f_\pi,\,m_K,\,f_K$), were computed. 
\item The concluding section contains comparisons with few other soliton model results and with 
experiments. Discussion about the SU(2)$_f$ versus SU(3)$_f$ Skyrme model considering 
symmetry breaking effects, the question of how different
methods and modified dynamical assumptions would lead to different results for the
nucleon axial-vector coupling, moments of inertia, absolute masses and mass splittings is given.
At the end we present our prediction for experimentally still missing, 
the $\overline{\bf 10}$ masses of penta-quark states $\rm N^*$ and $\Sigma_{\overline{10}}$,
for the whole ${\bf 27}_{3/2}$ absolute mass spectrum and
two smallest mass splittings among all of them
the $\Delta_3={\cal M}^{\bf 27}_{3/2} - {\cal M}^{\overline{\bf 10}}$ and 
the $\Delta_4={\cal M}^{\bf 35}_{5/2} - {\cal M}^{\bf 27}_{3/2}$.
\end{itemize}

\section{Minimal ${\rm SU(3)_f}$ extended Skyrme model}

\subsection{Basics of the Skyrme model}

By definition,  we introduce a theory with ${\rm SU(2)_{L} \times SU(2)_{R}}$ symmetry,
spontaneously broken into a diagonal SU(2) theory. Vacuum states of such a theory are in one-to-one
correspondence to SU(2), while low-energy dynamics is described by introducing a field U$(x_{\alpha})$
which has the property that, for every space-time point $x_{\alpha}$, the field U$(x_{\alpha})\in$  
SU(2), i.e. it is a $2\times 2$ matrix of determinant 1. Taking into account 
${\rm SU(2)_{L} \times SU(2)_{R}}$ (using matrices (A,B)), we can transform the field U into
$\rm U \rightarrow AUB^{-1}; \: A=U_{L}, \: B^{-1}=U^{\dagger}_{R}$.
The effective Lagrangian for U should have ${\rm SU(2)_{L} \times SU(2)_{R}}$ symmetry,
a possible minimal number of derivatives, and should 
correctly describe the low-energy limit: 
current algebra and partial conservation of axial currents (CA and PCAC) \cite{do2}.

The unique choice that satisfies the above conditions is the non-linear $\sigma$-model.
If the Lagrangian contains only the 
$\rm \partial_{\mu}  U\partial^{\mu}  U^{\dagger}$ term, the minimum energy in 
the sector with the soliton number $\not= 0$ is zero. 
This means that the soliton is reduced to the zero 
magnitude, i.e. it is unstable. 
To preserve the soliton from such reduction, i.e. to stabilize it, Skyrme
proposed \cite{sky} an additional 4-derivative term 
to the non-linear $\sigma$-model, so that
\begin{equation}
{\cal L} = {\cal L}_{\sigma} + {\cal L}_{Sk} =\int d^{4}x \left[\frac{f^2_{\pi}}{4} {\rm Tr} 
\left(\partial_{\mu} {\rm U}(x) \partial^{\mu} {\rm U}^{\dagger}(x)\right)+
\frac{1}{32e^{2}} {\rm Tr [(\partial_{\mu} U)U^{\dagger}(\partial^{\nu} U)U^{\dagger}]^2}\right].
\label{5}
\end{equation}
This is now a two-parameter theory with $e$  to be determined later on.
Using simple scaling argument it is easy to proof the above stability statement.

If ${\bf{\cal U}}$ is the soliton solution, then U$=\rm A{\bf{\cal U}}A^{\dagger}$ (for an arbitrary constant 
matrix A$\in $ SU(2)) is also a solution at the same finite energy as that of ${\bf{\cal U}}$,
but a solution for any A is not an eigenstate of spin and isospin. This leads to the so-called
null-frequency modes in the expansion around ${\bf{\cal U}}$.
The collective coordinate method \cite{wit}, which
treats A as a quantum-mechanical variable, eliminates these modes, and the Lagrangian and other
physical observables can be written as time-dependent functions ${\rm A}(t)$. 
The space-time dependent matrix field U$(\mathbf r,t) \, \in$ SU(2) takes the form:
\begin{equation}
{\rm U}(\mathbf r,t) = {\rm A}(t){\bf{\cal U}}(\mathbf r){\rm A}^{\dagger}(t),
\;\;
{\bf{\cal U}}(\mathbf r)= {\rm exp}(i{\mbox{\boldmath $\tau$}}{\cdot}{\mathbf r_{0}}F(r)),
\;\;
F(r) = 2{\rm arctan} \left[ \left({\frac{r_0}{r}}\right)^2 \right],
\label{22}
\end{equation}
with the famous SU(2)$_f$ Skyrme ansatz and $F(r)$ is the arctan ansatz for 
the profile function satisfying Euler-Lagrange equation \cite{dia,prat2}.
Here $r_0$ - the soliton size - is the variational parameter and the second power of $r_0/r$ is determined 
by the long-distance behavior of the equations of motion. After rescaling $x=ref_{\pi}$, we obtain the ratio
$r_0/r=x_0/x$. The quantity $x_0$ has the meaning of a dimensionless size of a soliton 
(or rather in units of $(ef_{\pi})^{-1}$). 
The advantage of using arctan ansatz is that all integrals involving
the profile function can be evaluated analytically. 
Hence, substitution of U$(\mathbf x,t) \, \in$ SU(2)$_f$ into ($\ref{5}$) gives 
well known classical result \cite{adk}
\begin{eqnarray}
{\cal L} &=& -{\cal M}_{csol}[F(x)] + \lambda[F(x)] {\rm Tr \dot A \dot A^{\dagger}} = 
-{\cal M}_{csol} + 2\lambda {\sum_{i=0}^{3}} {\dot a}^2_i, 
\label{28}
\end{eqnarray}
where ${\dot a}_i$ are angular velocities.

Here we have used the very well known group theoretical methods \cite{GROUP} 
the method to simplify the evaluation of the large and complicated terms.
This method is essentially an expansion of Lie-algebra elements 
$(\partial_{\mu} {\rm U}){\rm U}^{\dagger}$ over an adjoint representation of SU(N).
The coefficients of the expansion are known as the ``killing'' vectors.

Minimizing ${\cal M}_{csol}$ with respect to $x_0$, we get $x_0 = {\sqrt {15}}/4$. 
Then the classical mass and the moment of inertia for rotation 
in coordinate space $\lambda[F]$ reads \cite{prat2}:
\begin{eqnarray}
{\cal M}_{csol} = \frac{3}{2}{\sqrt {30}}{\pi}^2 \frac{f_{\pi}}{e}, 
\;\;
\lambda [F] = \frac{\pi}{3e^3 f_{\pi}} \frac{95}{32} {\sqrt {30}} \pi. 
\label{39}
\end{eqnarray}

By the prescription in \cite{adk} the soliton is quantized and the SU(2)$_f$ wave functions
were constructed. Next we use the variation equation and obtain the eigenenergies,
from which it follows \cite{adk,adk1} that
\begin{eqnarray}
m_N = {\cal M}_{csol} + \frac{3}{8\lambda} \: , \: m_{\Delta} = {\cal M}_{csol} + \frac{15}{8\lambda} \: , \: 
{\cal M}_{csol} = 81.09\frac{f_{\pi}}{e} \: , \: \lambda= \frac{51.08\pi}{3f_{\pi}e^3}. 
\label{40}
\end{eqnarray}
The model constants $f_{\pi}$ and $e$ are to be fixed, so that the masses $m_{N}$ and
  $m_{\Delta}$ should be reproduced. 
It has been found that $f_{\pi} = 64.5$ MeV and $e=5.45$ 
satisfies first statement within 8\% \cite{adk,adk1}.
For the physical values $f_{\pi} = 93$ MeV and $e=5.45$ we get ${\cal M}_{csol}=1384$ MeV. 
This value is too high, but nowadays 
nobody believes that absolute masses can be reproduced by the Skyrme model.
If one wants to use the physical value for $f_{\pi} = 93$ MeV, 
then it is necessary to choose $e = 4.825$ to reproduce the 
empirical mass difference $m_{\Delta} - m_N = 293$ MeV.

Next we approach the evaluation of the static properties of nucleons.
Using the arctan ansatz (\ref{22}) and performing the integration in $g_A$, we find 
the SU(2)$_f$ axial-vector coupling as a function of $x_0$:
\begin{eqnarray}
g_A (0)= \frac{-\pi}{3e^2}\left[-8x_0^2 - 4\pi\right]_{x_0 = \sqrt{15}/4}=\frac{\pi}{6e^2}(15+8\pi).
\label{44}
\end{eqnarray}
The integrals coming from the pure Lagrangian (\ref{5})
have logarithmic divergences of the same size and
of the $\Gamma(0)$ type, with opposite signs, so that they cancel each other, as they should. 
The Skyrme term stabilizes the soliton and does not 
create additional divergences in the calculation of $g_{\rm A}$.
This is an implicit proof of the necessity of adding the Skyrme
${\cal L}_{\rm Sk}$ term to the $\sigma$-model Lagrangian and that 
the arctan ansatz scheme, as a whole, works 
very well for the static properties of baryons.

Other static properties of nucleons, such as the isoscalar mean radius $R_I=\langle r^2\rangle^{1/2}_{I=0}$, 
the isoscalar magnetic mean radius $R_M=\langle r^2\rangle^{1/2}_{M,I=0}$, 
and proton/neutron magnetic moments in units of the nucleon Bohr magneton $\mu_B$, are:
\begin{eqnarray}
R^2_I \equiv \langle r^2\rangle_{I=0}\equiv 
\frac{\langle x^2\rangle_{I=0}}{e^2 f_{\pi}^2} 
=\frac{15}{4\pi e^2 f^2_{\pi}}, 
\;\;
R^2_M \equiv \langle r^2\rangle_{M,I=0}
=\frac{45\pi}{64e^2 f^2_{\pi}},
\;\;
{\mu}_{p \choose n} = \frac{m_{p \choose n}}{4f_{\pi}} \left[\frac{48e}{19\sqrt{30}\pi^3} \pm 
\frac{95\sqrt{30}\pi^2}{72e^3}\right]\; \mu_B.
\label{47}
\end{eqnarray}
For typical SU(2)$_f$ Skyrme model set of parameters, $f_{\pi}=64.5$ MeV, $e=5.45$, in the 
chiral limit 
\begin{eqnarray}
m_{\Delta}-m_N=293\; \rm MeV, \;\;g_A = 0.71,\;\;R_I = 0.61 \;\rm fm, \; \; 
R_M = 0.83 \;\rm fm, \;\;
\mu_p = 1.90\,\mu_B , \; \; \mu_n = -1.31\,\mu_B,
\label{48}
\end{eqnarray}
which are in nice agreement with the numerical evaluation of Ref. \cite{adk}.

\subsection{The ${\rm SU(3)_f}$ action, quantization and the construction of Noether currents}

Adding the Wess--Zumino term \cite{wess} and the minimal symmetry breaking term \cite{wei} to (\ref{5}), 
we obtain a chiral topological soliton model Lagrangian that describes baryons 
as topological excitations of a chiral effective
action depending only on meson fields. 
In the Introduction this model we have named the minimal SU(3)$_f$ extended Skyrme model ,
whose action is of the following form:
\begin{equation}
{\cal L} = {\cal L}_{\sigma} + {\cal L}_{\rm Sk} + {\cal L}_{\rm WZ} + {\cal L}_{\rm SB} ,
\label{49}
\end{equation}
\begin{equation}
{\cal L}_{\rm WZ} = \frac{-iN_{c}}{240 \pi^{2}} \int d\Sigma^{\mu \nu \rho \sigma \tau}
\rm Tr [U^{\dagger}\partial_{\mu}U \cdot U^{\dagger}\partial_{\nu}U \cdot U^{\dagger}\partial_{\rho}U
\cdot U^{\dagger}\partial_{\sigma}U \cdot U^{\dagger}\partial_{\tau}U],
\label{52}
\end{equation}
\begin{eqnarray}
{\cal L}_{\rm SB} 
 &=& \int d^{4}x \left\{ \frac{1}{24} \left(f^2_{\pi}m^2_{\pi}+2f^2_K m^2_K\right)\rm Tr [U+U^{\dagger} -2]
  + \frac{\sqrt{3}}{6} \left(f^2_{\pi}m^2_{\pi} - f^2_K m^2_K\right)\rm Tr [\lambda_{8}(U+U^{\dagger})]
	 \right.  \nonumber   \\ & &  \left.
  - \frac{1}{12} \left(f^2_{\pi} - f^2_K\right) \rm Tr [(1- \sqrt{3} \lambda_{8})
(U\partial_{\mu}U^{\dagger} \partial^{\mu} U 
+ U^{\dagger} \partial_{\mu}U\partial^{\mu}U^{\dagger})] \right\},
\label{53}
\end{eqnarray}
where  ${\cal L}_{\sigma}$, ${\cal L}_{\rm Sk}$, ${\cal L}_{\rm WZ}$ and ${\cal L}_{\rm SB}$ denote 
the $\sigma$-model, Skyrme, Wess--Zumino and symmetry breaking terms, respectively.
For U$(x)\in$ SU(2), the SB and WZ terms vanish.
The $f_{\pi(K)}$ and $m_{\pi(K)}$ are the pion (kaon) decay constants and masses, respectively.
Here the space-time dependent matrix field U$(\mathbf r,t) \, \in $ SU(3) takes the form:
\begin{equation}
{\rm U}(\mathbf r,t) = {\rm A}(t){\bf{\cal U}}(\mathbf r){\rm A}^{\dagger}(t),
\;\;\;
{\bf{\cal U}}(\mathbf r) = 
\left(
\begin{array}{c|c}
{\rm exp}(i{\mbox{\boldmath $\tau$}}{\cdot}{\mathbf r_{0}}F(r)) & 
{\begin{array}{c} 0 \\ 
0 \end{array}}
\\
\hline 
{\begin{array}{cc}
0 & 0
\end{array}}  &  1 
\end{array}\right).
\label{55}
\end{equation}
where $\rm {\bf{\cal U}}(\mathbf r)$ is the SU(3)$_f$ matrix to which the Skyrme SU(2)$_f$ ansatz is embedded.
The time-dependent collective coordinate matrix $\rm A(t)\, \in \, SU(3)$, introduced in (\ref{22}),
defines generalized (i.e. eight-angular) velocities $\dot a^{\alpha}$\cite{wit}:
\begin{equation}
{\rm A}^{\dagger} (t) \dot {\rm A}(t) = {i\over 2}\sum_{\alpha =1}^8 \lambda_{\alpha} \dot a^{\alpha}
= - {\dot {\rm A}}^{\dagger} (t) {\rm A}(t).
\label{56}
\end{equation}
Note that in addition to the general velocities $\dot a^{\alpha}$,
the adjoint matrix representation of the collective rotations ${\rm A}(t)$, 
\begin{eqnarray}
D_{\alpha}^{\beta} ({\rm A}) = \frac{1}{2} 
\rm Tr \left( \lambda_{\alpha} {\rm A}^{\dagger} \lambda^{\beta} {\rm A} \right) \; ; \; 
\alpha \, , \, \beta = 1,...,8 \: ; \: SU(3) \: \, {\rm indices},
\label{57}
\end{eqnarray}
will be important, especially for the case of flavor
symmetry breaking. 

In order to quantize the three-flavor Lagrangian (\ref{49}), we require that the spin and flavor 
operators should be the Noether charges. 
Owing to the structure of the Skyrme ansatz (\ref{55}), the infinitesimal change under 
spatial rotations could be expressed as a derivative with respect to $\dot{\vec{a}}$ \cite{wei}.

The symmetries of the quantized model, 
namely $\rm SU(3)_L-flavor \;\,and \;\,SU(3)_R-spin$, correspond, respectively,  
to the left (right) multiplication of A$(t)$ by a constant SU(3) [SU(2)] matrix. Therefore, the baryon 
wave functions are given as the matrix elements of the SU(3) representation functions \cite{gua}:
\begin{eqnarray}
\Psi ({\rm A}) = {\sqrt{{\rm dim}{\cal R}}}\;\langle YII_3|D^{\cal R} ({\rm A})|1S -S_3\rangle,
\label{59}
\end{eqnarray}
where $\cal R$ labels the SU(3) representation ($\cal R$ = 8,...); $Y,I,I_3$ stand for the 
hypercharge and the isospin, respectively, while $S, S_3$ denote the spin. 
The ``right'' hypercharge is constrained 
by the quantization condition $Y_R=N_c B/3=1$ for $N_c=3$ and $B=1$ \cite{gua}.
Namely only $Y_R \equiv 1$, following from Wess--Zumino action select the representations
of triality zero for $N_c=3$; i.e. it selects 
$\bf 8$, $\bf{10}$, $\bf{\overline{10}}$,  $\bf{27}$, $\bf{35}$, $\bf{\overline{35}}$, $\bf{64}$,...

After quantization and implementation of the above constraints, by a Legendre transformation,
the baryonic effective collective Hamiltonian was obtained from (\ref{49}). It has the
following eigenvalues \cite{gua}:
\begin{eqnarray}
M^{\cal R}_B=\underbrace{{\cal M}_{\rm sol}-{\cal M}}_{{\cal M}_{\rm csol}}+
\left(\frac{1}{2\lambda_c}-\frac{1}{2\lambda_s}\right)S(S+1)+\frac{1}{2\lambda_s}
\left(C_{2}({\cal R})-\frac{N_c^2}{12}\right)  -\frac{1}{2} \gamma\;\delta_{B}^{\cal R}.
\label{61}
\end{eqnarray}
Here $S$ denotes baryon spin, $C_{2}({\cal R})=(1/3)(p^2+q^2+pq+3(p+q))$ 
is the second order Casimir operator for an irreducible SU(3) representation ${\cal R}=(p,q)$.
The couplings $f_{\pi}$ and $f_K$ are no longer free parameters fitted to the absolute 
values of the baryon masses, but are 
real constants equal to its experimental values $f_{\pi}^{\rm exp} =93$ MeV and $f_K^{\rm exp}=113$ MeV.
This results from the SU(3)$_f$ extension of the Skyrme model, 
mainly by taking into account the Casimir operator of SU(3) and symmetry breaking effects. 

The classical soliton mass ${\cal M}_{\rm{sol}}[F]$, two moments
of inertia $\lambda_{c,s}[F]$ and symmetry breaker $\gamma [F]$ are functionals of the
solitonic solution $F(r)$.
The $\cal M$ is an unknown subtraction constant which takes into account
uncontrolled $1/N_{c}$ corrections. Therefore in principle the soliton mass
${\cal M}_{\text{csol}}$ can be treated as a free parameter.
The values of the SU(3) Casimir operator, spin $S$ for minimal and non minimal multiplets of baryons,
as a functions of an irreducible representation multiplets ${\cal R}=(p,q)$,
are given nicely in Table 1. of Ref. \cite{K}. Splitting constants $\delta_{B}^{\cal R}$,
will be defined later for specific representations $\cal R$.

Let us now construct the left (right) Noether currents associated with the $V-A$ $(V+A)$ 
transformations $(\delta U)_{\alpha} = i[Q_{\alpha},{\rm U}]; \; 
(\delta {\rm U}^{\dagger})_{\alpha} = i[Q_{\alpha},{\rm U}^{\dagger}]$ 
as a function of collective coordinates \cite{prat2}:
\begin{eqnarray}
J_{\mu\alpha}(L)& =& 
-\sum_{i,j} \left(\frac{\delta {\cal L}}{\delta \partial_{\mu} {\rm U}_{i,j}} (\delta {\rm U}_{i,j})_{\alpha} +
\frac{\delta {\cal L}}{\delta \partial_{\mu} {\rm U}^{\dagger}_{i,j}}
(\delta {\rm U}^{\dagger}_{i,j})_{\alpha}\right)= 
J_{\mu\alpha}^{\sigma}(L) + J_{\mu\alpha}^{\rm Sk}(L) 
                    + J_{\mu\alpha}^{\rm SB}(L) + J_{\mu\alpha}^{\rm WZ}(L) \, ,
\label{63}  \\
J_{\mu\alpha}^{\sigma}(L)& =& \frac{i}{2}f_{\pi}^2\rm Tr \left\{ Q_{\alpha}(\partial_{\mu}U)
                    U^{\dagger}\right\}\,,\;\;
J_{\mu\alpha}^{\rm Sk} (L) = \frac{i}{8e^2} \rm Tr \left\{ [(\partial_{\nu}U) U^{\dagger},Q_{\alpha}] 
                  [(\partial_{\mu} U)U^{\dagger},(\partial^{\nu} U)U^{\dagger}] \right\}
	     \, , \nonumber \\
J_{\mu\alpha}^{\rm WZ}(L)& =& \frac{-N_c}{48\pi^2} \epsilon_{\mu \nu \rho \sigma}
         \rm Tr \left\{Q_{\alpha}(\partial^{\nu}U)(U^{\dagger}
         \partial^{\rho}U)(U^{\dagger}\partial^{\sigma}U)U^{\dagger}\right\}, \;
J_{\mu\alpha}^{\rm SB} (L)= \frac{f_{\pi}^2 - f^2_K}{12i}\rm Tr \left[\left(1-\sqrt 3 \lambda_8 \right)
\left[U,Q_{\alpha}\right] (\partial_{\mu}U)U^{\dagger} \right] \, , 
	 \nonumber
\end{eqnarray}
where the superscripts $\sigma$, Sk, WZ stand for the $\sigma$-model, Skyrme and Wess--Zumino currents,
reflecting the fact that the currents (\ref{63}) come from different pieces of the Lagrangian (\ref{49}).
Specifying the the Noether charge matrices $Q_{\alpha}$,
we obtain the SU(2)$_f$ or SU(3)$_f$ currents, respectively. From 
$J_{\mu}^{(V+A)} = J_{\mu}^{(V-A)}({\rm U} \, \leftrightarrow \, {\rm U}^{\dagger})$
it is simple to find $J_{\mu}^{(V,A)}$.

\begin{figure}[h]
\centerline{\includegraphics{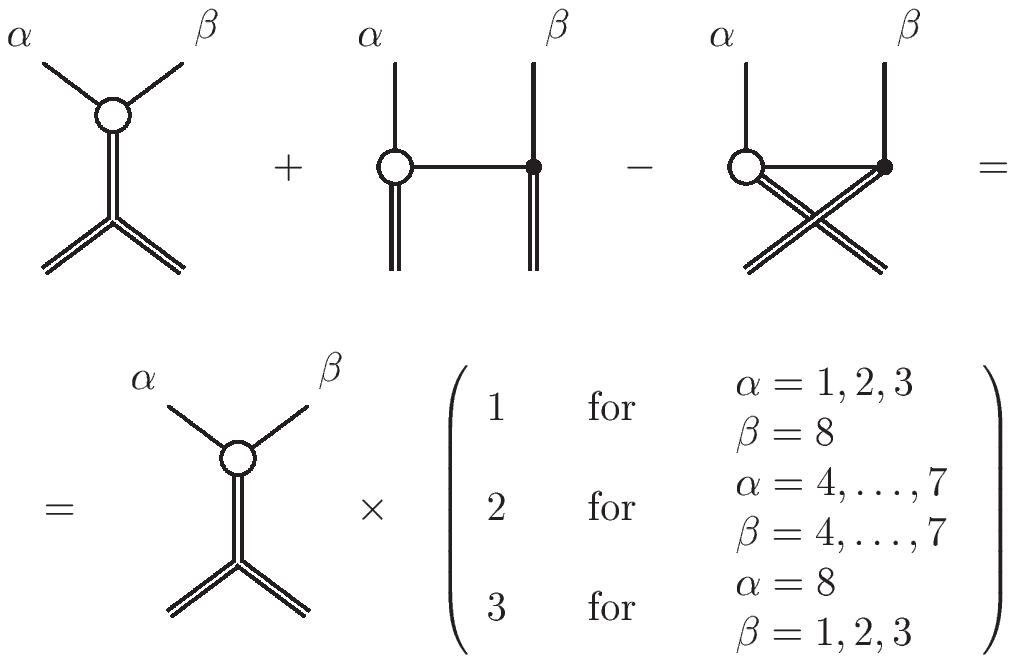}}
\caption{Graphical computation of the Wess--Zumino current quantity $N(\tau)$}
\label{fig1}
\end{figure}

Inserting the space-time dependent matrix field ${\rm U}({\mathbf r},t)$ from Eq. (\ref{55}) into the relations 
(\ref{63}) and applying the ``killing'' vector method,
we obtain the following time and space components of the $\sigma$-model, Skyrme and Wess--Zumino currents:
\begin{eqnarray}
J_{0\alpha}^{\sigma}(L) &= &\frac{i}{2}f_{\pi}^2 D_{\alpha}^{\beta}(A)
       \left\{ \delta_{\beta}^{\gamma} - {\cal U}_{\beta}^{\gamma} \right\} 
       {\dot a}^{\gamma} \, , \;\;
J_{i\alpha}^{\sigma}(L) = \frac{i}{2}f_{\pi}^2 D_{\alpha}^B(A) {\xi}_i^B
             \, ,\label{71} \\
J_{0\alpha}^{\rm Sk}(L) &= & \frac{i}{8e^2} D_{\alpha}^{\beta}(A)
        {\xi}^A_j {\xi}^B_j \left\{ {\delta}^{\gamma}_{\beta} 
	-  {\cal U}_{\beta}^{\gamma} \right\}
        f_{A\gamma\tau}f_{\tau\rho B}{\dot a}^{\rho} \, , \nonumber \\
J_{i\alpha}^{\rm Sk}(L) &=& \frac{i}{8e^2} \left\{ {D_{\alpha}^B}(A) {\xi}^A_j 
      ( {\xi}^A_j {\xi}^B_j -{\xi}^B_j {\xi}^A_i ) 
       + D_{\alpha}^{\rho} (A) {\cal U}_{\rho}^{\tau} {\xi}^A_i 
       f_{\beta A \sigma}f_{\sigma\tau\gamma}
   ({\dot a_i} {\cal U}_i^{\beta} -{\dot a}^{\beta})
  ({\dot a_i} {\cal U}_i^{\gamma}- {\dot a}^{\gamma}) \right\}
           \, ,\label{73} \\
J_{0\alpha}^{\rm WZ}(L) &=& \frac{N_c}{192 \pi^2} D_{\alpha}^{\beta}(A) 
      d_{AD\beta} \epsilon^{ijk} \epsilon_{DBC} \eta^A_i\eta^B_j \eta^C_k, 
      \nonumber\\
J_{i\alpha}^{\rm WZ}(L) &=& \frac{-N_c}{192 \pi^2} D_{\alpha}^{\beta}(A) 
     \epsilon^{ijk} \left\{ \delta^{\tau}_{\beta} \eta^A_j \eta^B_k 
     - {\cal U}^{\tau}_{\beta} \xi^A_j \xi^B_k
     \right\} N(\tau) f_{ABC} d_{ C \tau \gamma} {\dot a}^\gamma,
     \label{75} 
\end{eqnarray}
where $i, j = 1,...,3$ are the Euclidean space indices; $A, B, C = 1,...,3$ are the isospin
SU(3) indices. The above currents contain the following definitions:
\begin{eqnarray}
(\partial_i {\bf{\cal U}}){\bf{\cal U}}^{\dagger} = \frac{i}{2} \lambda_{\alpha} \xi^{\alpha}_i \, ; \;
{\bf{\cal U}}^{\dagger} \lambda_{\alpha} {\bf{\cal U}} = \lambda_{\beta} {\bf{\cal U}}_{\alpha}^{\beta}\, ;\;
{\bf{\cal U}}^{\dagger}(\partial_i {\bf{\cal U}}) = \frac{i}{2} \lambda_{\alpha} \eta^{\alpha}_i \, ; \;
{\bf{\cal U}} \lambda_{\alpha} {\bf{\cal U}}^{\dagger} = \lambda_{\beta} {\bf{\cal V}}_{\alpha}^{\beta}\, , 
\label{76}
\end{eqnarray}
where ($\xi_i^{\alpha},\:{\bf{\cal U}_{\alpha}^{\beta}}$) and ($\eta_i^{\alpha},\:{\bf{\cal V}_{\alpha}^{\beta}}$) 
are the so-called left and right SU(3)
``killing'' vector components, respectively. They have the following properties:
\begin{eqnarray}
\xi_i = {\bf{\cal U}}\eta_i \; ,\; \eta_i = {\bf{\cal V}}\xi_i \: \: \iff \; \; {\bf{\cal U}} = {\bf{\cal V}}^{-1}\,.
\label{77}
\end{eqnarray}
The SU(3) Wess--Zumino current quantity $N(\tau)$ 
is evaluated with the help of the group theory for Feynman diagrams in 
non-Abelian gauge theories \cite{cv}. Using this method with group-theoretical identities,
such as the Lie commutators, we obtain a graphic expression, Figure \ref{fig1},
from which we easily obtain the desired Wess--Zumino current quantity:
\begin{eqnarray}
N({\tau}) = \cases{1, & $\tau = 1,...,3$,\cr 
2, & $\tau = 4,...,7$,\cr
3, & $\tau = 8$.\cr}
\label{78}
\end{eqnarray}
For the minimal SU(3)$_f$ extended Skyrme model polar components of the ``killing'' vectors 
$\xi_i^{\alpha} \, , \, \eta_i^{\alpha}$, and ${\bf{\cal U}}_{\alpha}^{\beta}$ are 
computed and presented in tabular forms, \ref{t:tab1}--\ref{t:tab3}. 
\renewcommand{\arraystretch}{1.5}
\begin{table}
\caption{``Killing'' vectors ${\xi}_i^j$ and ${\eta}_i^j$  in polar coordinates}
\begin{center}
\begin{tabular}{|c|c|c|c|}
\hline 
$ i $ & $ r $ & $ \theta $ & $ \phi $ \\
\hline \hline
$\xi^1_i$ & $ 2 \, F' \, \sin{\theta} \, \cos{\phi}$ & $ \sin2F \, \cos{\theta} \, \cos{\phi} +$ & 
$ -\sin2F \, \sin{\theta} \, \sin{\phi} +$ \\
  & $$ & $ 2\sin^2 F \, \sin{\phi} $ & 
  $ \sin^2 F \, \sin2{\theta} \, \cos{\phi} $ \\
\hline
$\xi^2_i$ & $ 2F' \, \sin{\theta} \, \sin{\phi} $ & $ \sin^2 F \, \cos{\theta} \, \sin{\phi} - $ & 
$ \sin2F \, \sin{\theta} \, \cos{\phi} +$ \\
  &  & $ 2\sin^2 F \, \cos{\phi} $ & $ \sin^2 F \, \sin2{\theta} \, \sin{\phi} $ \\
\hline
$\xi^3_i$ & $ 2 \, F' \, \cos{\theta}$ & $ -\sin2F \, \sin{\theta}$ &
$ -2 \sin^2 F \, \sin^2{\theta}$ \\
\hline 
$\eta^1_i$ & $ 2 \, F' \, \sin{\theta} \, \cos{\phi}$ & $ \sin2F \, \cos{\theta} \, \cos{\phi} -$ & 
$ -\sin2F \, \sin{\theta} \, \sin{\phi} -$ \\
  & $$ & $ 2\sin^2 F \, \sin{\phi} $ & 
  $ \sin^2 F \, \sin2{\theta} \, \cos{\phi} $ \\
\hline
$\eta^2_i$ & $ 2F' \, \sin{\theta} \, \sin{\phi} $ & $ \sin^2 F \, \cos{\theta} \, \sin{\phi} + $ & 
$ \sin2F \, \sin{\theta} \, \cos{\phi} -$ \\
  &  & $ 2\sin^2 F \, \cos{\phi} $ & $ \sin^2 F \, \sin2{\theta} \, \sin{\phi} $ \\
\hline
 $\eta^3_i$ & $ 2 \, F' \, \cos{\theta}$ & $ -\sin2F \, \sin{\theta}$ &
 $ 2 \sin^2 F \, \sin^2{\theta} $ \\
\hline 
\end{tabular}\\
${\xi}_i^A = -i {\rm Tr} ((\partial_i {\cal U}){\cal U}^{\dagger}\lambda^A) \: ;  \: 
{\eta}_i^A = -i {\rm Tr} ({\cal U}^{\dagger}(\partial_i {\cal U})\lambda^A) \: ;$ \\
$A=1,2,3,8 \: ; \: {\xi}_i^8 = {\eta}_i^8 = 0$
\end{center}
\label{t:tab1}
\end{table}
\begin{table}
\caption{``Killing'' vectors ${\cal U}_{\alpha}^{\beta} = 
\frac{1}{2} \rm Tr (\lambda_{\alpha}{\cal U}^{\dagger} \lambda^{\beta}{\cal U})$ in polar coordinates}
\begin{center}
\begin{tabular}{|c|c|c|c|}
\hline
$ \beta $ & $ 1 $ & $ 2 $ & $ 3 $ \\
\hline \hline
${\cal U}_1^{\beta}$ & $ \cos^2 F - \sin^2 F \cos^2{\theta} + $ & $ \sin2F \cos{\theta} +$ & 
$ -\sin2F \, \sin{\theta} \, \sin{\phi} +$ \\
  & $ \sin^2 F \, \sin^2{\theta} \, \cos2{\phi} $ & $ \sin^2 F \, \sin^2{\theta} \, \sin2{\phi} $ & 
  $ 2\sin^2 F \, \sin2{\theta} \, \cos{\phi} $ \\
\hline
${\cal U}_2^{\beta}$ & $ -\sin2F \, \cos{\theta} + $ & $ \cos^2 F - \sin^2 F \cos^2{\theta} - $ & 
$ \sin2F \, \sin{\theta} \, \cos{\phi} +$ \\
  & $ \sin^2 F \, \sin^2{\theta} \, \sin2{\phi} $ & $ -\sin^2 F \, \sin^2{\theta} \, \cos2{\phi} -$ & 
$ \sin^2 F \, \sin2{\theta} \, \sin{\phi} $ \\
\hline
${\cal U}_3^{\beta}$ & $ \sin2F \, \sin{\theta} \, \sin{\phi} +$ & $ -\sin2F \, \sin{\theta} \, \cos{\phi}+$ &
$ 1-2 \sin^2 F \, \sin^2{\theta}$ \\
  & $ \sin^2 F \, \sin2{\theta} \, \cos{\phi} $ & $ \sin^2 F \, \sin2{\theta} \, \sin{\phi} $ & \\
\hline
\end{tabular}\\
{Other components are:} \, 
${\cal U}_{\alpha}^{\beta} = 0 \; \rm for \; \alpha = 1,2,3 \; ; \; \beta = 4,5,6,7,8 $
\end{center}
\label{t:tab2}
\end{table}
\begin{table}
\caption{``Killing'' vectors ${\cal U}_{\alpha}^{\beta} = 
\frac{1}{2} \rm Tr (\lambda_{\alpha}{\cal U}^{\dagger} \lambda^{\beta}{\cal U})$ in polar coordinates}
\begin{center}
\begin{tabular}{|c|c|c|c|c|}
\hline
$ \beta $ & $ 4 $ & $ 5 $ & $ 6 $ & $ 7 $ \\
\hline \hline
${\cal U}_4^{\beta}$ & $ \cos F $ & $ \sin F \cos{\theta} $ & $ \sin F \sin{\theta} \sin{\phi}$ &
$ \sin F \, \sin{\theta} \, \cos{\phi} $ \\
\hline
${\cal U}_5^{\beta}$ & $ -\sin F \, \cos{\theta} $ & $ \cos F $ & $ -\sin F \, \sin{\theta} \, \cos{\phi} $ &
$ \sin F \sin{\theta} \sin{\phi}$ \\
\hline
${\cal U}_6^{\beta}$ & $ -\sin F \sin{\theta} \sin{\phi} $ & $ \sin F \, \sin{\theta} \, \cos{\phi} $ & 
$ \cos F $ & $ -\sin F \cos{\theta} $\\
\hline
${\cal U}_7^{\beta}$ & $ -\sin F \sin{\theta} \cos{\phi} $ & $ -\sin F \, \sin{\theta} \, \sin{\phi} $ & 
$ \sin F \cos{\theta} $ & $ \cos{\theta} $\\
\hline
\end{tabular}\\
{Other components are:} ${\cal U}_{\alpha}^{\beta} = 0 \; \rm for \; \alpha = 4,5,6,7,8 \; ; \; 
\beta = 1,2,3 \;  \& \;  {\cal U}_8^8 = 1$
\end{center}
\label{t:tab3}
\end{table}
\renewcommand{\arraystretch}{1}

\subsection{Arctan ansatz as the ${\rm SU(3)_f}$ profile function F(r) and dynamics 
of the SU(3)$_f$ symmetry breaking}

For the SU(3)$_f$ extension (\ref{49}) of the Skyrme Lagrangian (\ref{5}), 
we use a new set of parameters ${\hat x}, 
{\beta}', {\delta}'$ introduced in Ref. \cite{wei}.
The symmetry breaker ${\hat x}$ was constructed systematically from the QCD mass term.
The $\delta'$ term is required to split pseudoscalar meson masses, while the $\beta'$ term is required
to split pseudoscalar decay constants \cite{wei}:
\begin{eqnarray}
\beta' =\frac{f^2_K - f^2_{\pi}}{4(1-{\hat x})},
\; \; \delta' = \frac{m^2_{\pi}f^2_{\pi}}{4} = \frac{m^2_K f^2_K}{2(1+{\hat x})}, \; \; 
{\hat x}=\frac{2m^2_K f^2_K}{m^2_{\pi}f^2_{\pi}} - 1.
\label{82}
\end{eqnarray}
Considering the above symmetry breaking parameters we are introducing three 
different dynamical assumptions, based on the SB (\ref{53}), producing three
fits which are going to be used further on in our numerical analysis:
\begin{eqnarray}
&({\rm i})&\;\hspace{0.2mm} m_\pi=m_K=0, \;\,f_\pi=f_K=93\;\, {\rm MeV}\; 
\Longrightarrow \;\, {\hat x}=1,\;\, \beta' = \delta' = 0;
\label{83}\\
&({\rm ii})&\;\hspace{0.1mm} m_\pi=138,\;\, m_K=495, \;\,f_\pi=f_K=93,\,\; {\rm MeV}\; \Longrightarrow\, 
\hspace{5mm}{\hat x}=24.73,\,\; \beta' =0,\,\;\delta' = 4.12\,\times 10^7,\,{\rm MeV}^4;
\nonumber\\
&({\rm iii})& \,m_\pi=138,\,\; m_K=495,\,\;f_{\pi}=93,\,\;f_K =113\, {\rm MeV} \Longrightarrow 
{\hat x}=36.97,\;\, \beta' = -28.6\;\,{\rm MeV}^2,\,\;\delta' = 4.12\,\times 10^7,\,{\rm MeV}^4.
\nonumber
\end{eqnarray}
Fit (i) corresponds to the SU(3)$_f$ chiral limit.
For the numerical results, which are going to be presented in Tables \ref{t:tab4}-\ref{t:tab9}, 
we choose typical range of the Skyrme charge values $3.0\;\leq\;e\;\leq \;5.0$. 
The reason for this lies in the fact that 
in the most realistic case (iii), $e=3.4,\;4.2,\;4.4,\;4.7$, 
gives the best fit for axial-vector coupling $g_A$,
the octet-decuplet mass splitting $\Delta_1$,  
and for the penta-quark masses $M_{\Theta^+}$ and $M_{\Xi^{--}_{3/2}}$, respectively. 

Substituting (\ref{55}) into (\ref{49}) we obtain classical soliton mass
${\cal M}_{\rm csol}$ containing the symmetry breakers $\hat x$, $\beta'$ and $\delta'$.
Owing to their presence in ${\cal M}_{\rm csol}$ the dimensionless size of skyrmion $x_0$ 
is affected, i.e. instead of being a constant it becomes a complicated  
function of $e$, $m_{\pi}$, $f_{\pi}$, $m_K$, $f_K$, or via Eqs. (\ref{82}) 
a function of $e$, $f_{\pi}$, $\beta'$, and $\delta'$:
$x_0 \longrightarrow x_0^{SB} = x_0^{SB} (e,f_{\pi},\beta',\delta') \equiv x'_0$. 
The analytical expression for the SU(3)$_f$ extended classical soliton mass:
\begin{eqnarray}
{\cal M}_{\rm csol}[F] &=& 2\pi \frac{f_{\pi}}{e} \int_0^{\infty} dx \left\{\left(x^2 {F'}^2 + 2 \sin^2 F \right) + 
\sin^2 F \left( 2{F'}^2 + \frac{\sin^2 F}{x^2} \right) 
\right. \nonumber \\  & & \left.
+ \frac{8}{f^2_{\pi}}(1-\cos F)\left[{\beta}'\left(x^2 {F'}^2 + 2 \sin^2 F \right) +
\frac{{\delta}'}{e^2f^2_{\pi}} x^2 \right]\right\} \nonumber \\
&=& 3{\sqrt 2}{\pi}^2 \left[x'_0 + \frac{15}{16x'_0} + \frac{2}{f^2_{\pi}}\left(3{\beta}'x'_0 +
\frac{4}{3} \frac{\delta'}{e^2 f^2_{\pi}} x'^3_0 \right)\right]\frac{f_{\pi}}{e},
\label{85}  
\end{eqnarray}
we are using next to obtain dimensionless size of skyrmion $x'_0$. 
Minimizing ${\cal M}_{\rm csol}$ with respect to $x'_0$, we have found : 
\begin{eqnarray}
{x^{\prime}_0}^2 = \frac{15}{8} \left[ 1 + \frac{6\beta'}{f^2_{\pi}} + 
\sqrt {\left(1+ \frac{6\beta'}{f^2_{\pi}}\right)^2 + 
\frac{30\delta'}{e^2 f^4_{\pi}}} \; \right]^{-1},
\label{86} 
\end{eqnarray}
which analytically describes  
the dynamics of skyrmion internal SU(3)$_f$ symmetry breaking effects.
This is our main result and it is clear from the above equation that a
skyrmion effectively shrinks when the Skyrme charge $e$ receives smaller values 
and it shrinks more when one ``switches on'' symmetry breaking effects.
The influence of the internal skyrmion dynamics, due to the symmetry breaking effects, 
on the nucleon axial-vector
current, the mass spectrum and the mass differences are going to be presented in tabular form.

\subsection{Nucleon axial-vector coupling, moments of inertia,  
the {\bf 8}, {\bf 10}, $\overline{\bf 10}$, ${\bf 27}_{3/2}$ mass spectrums and \\
the higher SU(3)$_f$ representation mass splittings}

Including the previously introduced arctan ansatz for the profile function $F(r)$,
in the minimal SU(3)$_f$ extended Skyrme model currents (\ref{63}),
we first compute the analytical expressions for the nucleon axial-vector coupling constant $g_A(x^{\prime}_0)$,
the moment of inertia for rotation in coordinate space $\lambda_c(x^{\prime}_0)$,
the moment of inertia for flavor rotations in the direction of the strange degrees of freedom,
except for the eight directions $\lambda_s(x^{\prime}_0)$ and the SU(3)$_f$ 
symmetry breaking quantity $\gamma(x^{\prime}_0)$ relevant to evaluate 
the higher SU(3)$_f$ representation mass splittings 
as well as the {\bf 8}, {\bf 10}, $\overline{\bf 10}$ and ${\bf 27}_{3/2}$ absolute 
mass spectrum \cite{wei,prat2}, respectively.
\begin{eqnarray}
g_A [F]&=&\frac{\pi}{e^2}\int_0^{\infty}dx
\left\{\left[ F'\left(x^2+2\sin^2F\right)
+ x\sin2F\left(1+{F'}^2\right)+\frac{\sin^2F}{x}\sin2F\right] \frac{-7}{30}
\right. \nonumber \\  & & \left.
+ \frac{4{\beta}'}{3f^2_{\pi}}\left(1-\hat x\right)(1-\cos F)
\left[\left(\frac{2\sin F}{x}(1+2\cos F)+F'\right)\frac{-49}{300}
+\left(\frac{2\sin F}{x}-F'\right)\frac{-1}{75}\right]
\right. \nonumber \\  & & \left.
+ \frac{eN_c}{12{\pi}^2 f_{\pi}\lambda_s [F]} (1-\cos F)\frac{\sin F}{x}
\left(\frac{\sin F}{x}-2F'\right)\frac{14}{30}\right\} \nonumber \\
&=&  \frac{14\pi}{15e^2}\left(2x'^2_0 + \pi\right) + 
(1-{\hat x})\frac{16\pi\beta'}{225 e^2 f^2_{\pi}} {x'}_0^2 + \frac{7\sqrt{2}N_c}{192ef_{\pi}}
\frac{x_0^{\prime}}{\lambda_s(x_0^{\prime})},
\label{87} \\
\lambda_c[F] &=& \frac{8\pi}{3e^3 f_{\pi}} \int_0^{\infty} dx 
\sin^2 F \left\{ x^2 \left( 1 + {F'}^2 + \frac{8{\beta}'}{f^2_{\pi}}(1-\cos F) \right)+ \sin^2 F \right\} \nonumber \\
&=&\frac{\sqrt 2 \pi^2}{3e^3 f_{\pi}} \left[6\left( 1 + 
2\frac{\beta'}{f^2_{\pi}} \right)x'^3_0 + \frac{25}{4} x'_0\right],
\label{88} \\
\lambda_s[F] &=& \frac{\pi}{2e^3 f_{\pi}} \int_0^{\infty} dx (1-\cos F)
\left\{x^2\left[4+{F'}^2 -\frac{16{\beta}'}{f^2_{\pi}}\left({\hat x}+\cos F\right)\right]+2\sin^2 F\right\}\nonumber\\
&=&\frac{\sqrt 2 \pi^2}{4e^3 f_{\pi}} \left[4\left( 1 - 
2\left(1+2{\hat x}\right)\frac{\beta'}{f^2_{\pi}} \right){x'}^3_0 + \frac{9}{4} x'_0\right],
\label{89} \\
\gamma [F] &=& \frac{32\pi}{3}\frac{{\hat x} -1}{e^3 f^3_{\pi}} \int_0^{\infty} dx 
\left\{ {\delta}'x^2 (1-\cos F) -
 e^2 f^2_{\pi} {\beta}' \cos F \left( x^2{F'}^2 + 2\sin^2 F \right)\right\}\nonumber \\
&=& 4 \sqrt 2 {\pi}^2 \frac{1-{\hat x}}{ef_{\pi}} \left[{\beta}'x'_0 -
\frac{4}{3} \frac{\delta'}{e^2 f^2_{\pi}} x'^3_0 \right] .
\label{90} 
\end{eqnarray}

\renewcommand{\arraystretch}{1.1}
\begin{table}
\caption{The SU(3)$_f$ Skyrme model dimensionless size of skyrmion $ x'_0 $,
axial-vector coupling constant $g_A$, the rotational moments of inertia $\lambda_{c(s)}$ (GeV$^{-1}$)
and the SB quantity $\gamma$ (GeV), for fits (i), (ii) and (iii). 
For fit (i) $x^{\prime}_0 \equiv x_0 =\frac{\sqrt {15}}{4}$ and $\gamma =0$.}
\begin{center}
\begin{tabular}{|c|ccc|ccccc|ccccc|}
\hline
${\rm Fits}$ & $ $ & $({\rm i})$ & $ $ & $ $ & $ $ & $({\rm ii})$ & $ $ & $ $ & $ $ & $ $ & $ ({\rm iii})$ & $ $ & $ $\\
\hline \hline
$ e $ & $g_A$ & $\lambda_c$ & $\lambda_s$ & $ x_0^{\prime}$ & $g_A$ & $\lambda_c$ & $\lambda_s$ & $\gamma$ 
& $ x_0^{\prime}$ & $g_A$ & $\lambda_c$ & $\lambda_s$ & $ \gamma $ \\
\hline 
$3.0$ & $ 1.701 $ & $ 21.304 $ & $ 8.073 $ & $ 0.8359 $ & $ 1.558 $ & $ 16.172 $ & $ 5.860 $ & $ 1.956 $ &
$ 0.8408 $ & $ 1.548 $ & $ 16.300 $ & $ 7.572 $ & $ 3.192 $\\
 
$3.2$ & $ 1.512 $ & $ 17.554 $ & $ 6.652 $ & $ 0.8465 $ & $ 1.399 $ & $ 13.635 $ & $ 4.960 $ & $ 1.674 $ &
$ 0.8517 $ & $ 1.386 $ & $ 13.750 $ & $ 6.429 $ & $ 2.751 $\\
 
$3.4$ & $ 1.358 $ & $ 14.635 $ & $ 5.546 $ & $ 0.8561 $ & $ 1.268 $ & $ 11.602 $ & $ 4.235 $ & $ 1.444 $ &
$ 0.8615 $ & $ 1.251 $ & $ 11.705 $ & $ 5.504 $ & $ 2.388 $\\
 
$3.6$ & $ 1.231 $ & $ 12.329 $ & $ 4.672 $ & $ 0.8646 $ & $ 1.159 $ & $  9.953 $ & $ 3.644 $ & $ 1.253 $ &
$ 0.8703 $ & $ 1.140 $ & $ 10.046 $ & $ 4.748 $ & $ 2.088 $\\

$3.8$ & $ 1.125 $ & $ 10.483 $ & $ 3.972 $ & $ 0.8723 $ & $ 1.068 $ & $  8.602 $ & $ 3.158 $ & $ 1.094 $ &
$ 0.8782 $ & $ 1.046 $ & $ 8.685 $ & $ 4.123 $ & $ 1.836 $\\
 
$4.0$ & $ 1.038 $ & $ 8.988 $ & $ 3.406 $ & $ 0.8792 $ & $ 0.992 $ & $  7.483 $ & $ 2.754 $ & $ 0.960 $ &
$ 0.8854 $ & $ 0.967 $ & $ 7.559 $ & $ 3.603 $ & $ 1.624 $\\

$4.1$ & $ 0.999 $ & $ 8.346 $ & $ 3.163 $ & $ 0.8824 $ & $ 0.959 $ & $  6.996 $ & $ 2.577 $ & $ 0.902 $ &
$ 0.8887 $ & $ 0.932 $ & $ 7.068 $ & $ 3.375 $ & $ 1.531 $\\
 
$4.2$ & $ 0.964 $ & $ 7.764 $ & $ 2.942 $ & $ 0.8855 $ & $ 0.928 $ & $  6.550 $ & $ 2.415 $ & $ 0.847 $ &
$ 0.8918 $ & $ 0.899 $ & $ 6.618 $ & $ 3.166 $ & $ 1.444 $\\

$4.4$ & $ 0.903 $ & $ 6.753 $ & $ 2.559 $ & $ 0.8911 $ & $ 0.874 $ & $  5.764 $ & $ 2.130 $ & $ 0.751 $ &
$ 0.8976 $ & $ 0.842 $ & $ 5.827 $ & $ 2.796 $ & $ 1.291 $\\

$4.6$ & $ 0.851 $ & $ 5.910 $ & $ 2.239 $ & $ 0.8962 $ & $ 0.830 $ & $  5.099 $ & $ 1.887 $ & $ 0.669 $ &
$ 0.9029 $ & $ 0.794 $ & $ 5.156 $ & $ 2.481 $ & $ 1.159 $\\

$4.8$ & $ 0.809 $ & $ 5.201 $ & $ 1.971 $ & $ 0.9009 $ & $ 0.792 $ & $  4.532 $ & $ 1.680 $ & $ 0.598 $ &
$ 0.9076 $ & $ 0.753 $ & $ 4.583 $ & $ 2.212 $ & $ 1.045 $\\

$5.0$ & $ 0.773 $ & $ 4.602 $ & $ 1.744 $ & $ 0.9051 $ & $ 0.761 $ & $  4.045 $ & $ 1.502 $ & $ 0.537 $ &
$ 0.9122 $ & $ 0.718 $ & $ 4.092 $ & $ 1.979 $ & $ 0.946 $\\
\hline \hline 
$ \rm Exp.$ & $ 1.26 $ & $ - $ & $ - $ &$ - $ & $ 1.26 $ & $ - $ & $ - $ & $ - $ & $ - $ &
$ 1.26 $ & $ - $ & $ - $ & $ - $\\
\hline \hline 
\end{tabular}
\label{t:tab4}
\end{center}
\end{table}

The last remaining quantity, $\gamma$, is an important coefficient in the symmetry breaking 
piece ${\cal L}_{\rm SB} = -\frac{1}{2} \gamma (1-D_{88})$ of a total collective Hamiltonian (\ref{61}) 
and is linear in the symmetry breaking parameter $(1-{\hat x})$.
Switching off SU(3)$_f$ symmetry breaking and in the chiral limit 
$\beta'=\delta'=0$, $x_0^{\prime} \longrightarrow x_0=\sqrt{15}/4$,
the $\lambda_c/\lambda_s$ becomes 95/36 and $\gamma =0$.
For $e=4$, all above expressions are in very good agreement with the values (2.48)
from Ref. \cite{wei} where fine tuning effects, like vector-meson contributions, 
the so-called static fluctuation, vibrations of Kaons, etc., \cite{wei}
are taken into account. 
For example, $\beta^2 \equiv \lambda_s (x_0^{\prime}) = 3.62$ GeV$^{-1}$ is very close
to the value of $3.52$ GeV$^{-1}$ mentioned in the discussion below Eq. (2.48) on p. 2440 of Ref. \cite{wei}.
This represents an implicit proof that the inclusion of 
the fine tuning effects does not change our results dramatically
and is one of our main reasons to concentrate on the axial-vector coupling and moments of inertia only. 
Their numerical values as a functions of $e$ are given in Table \ref{t:tab4}, 
while graphical displays are given for cases (i), (ii) and (iii) in Figures \ref{fig2}, \ref{fig3}
and \ref{fig4}, respectively. 

Moments of inertia we need to predict the {\bf 8}, {\bf 10}, $\overline{\bf 10}$ and 
{\bf 27} mass spectrums
and the higher SU(3)$_f$ representation mass splittings, 
while the evaluation of nucleon axial-vector coupling $g_A$ 
in this paper serves only as a consistency check of the approach as a whole.
Inspection of our Table \ref{t:tab4} case (iii), shows that $g_A=0.97$ for $e=4$, 
agrees within 1\% with value $g_A=0.98$ from page 2449 in Ref. \cite{wei}.
Other quantities, like magnetic moments and charge radii, 
for the minimal SU(3)$_f$ extended Skyrme model, behave in the same way 
and it is not necessary to present them here. For them we simply refer to the complete calculation 
presented in Table 2.2 of Ref. \cite{wei}.
However, for the sake of comparison with the SU(2)$_f$ results (\ref{44}-\ref{48}), and because we are also interested
in the influence of the SB dynamics (\ref{86}) on the leading 
SU(2)$_f$ terms of charge radii and magnetic moments of nucleons, 
we next estimate the isoscalar and the isovector components of magnetic moments of proton and neutron,  
nucleon isoscalar mean radius $R_I$,  and isoscalar magnetic mean radius $R_M$,
and present them as functions of $e$ and for fits (\ref{83}), in Table \ref{t:tab5}.
\begin{eqnarray}
{\mu}_{p \choose n} [F] &=& \frac{m_{p \choose n}}{3} 
\left[\frac{2{x'}_0^2}{\pi e^2 f^2_{\pi}\lambda_c (x_0^{\prime})} \pm \lambda_c (x_0^{\prime})\right], 
\label{91} \\
R^2_I [F]&=& \frac{-2}{\pi e^2 f_{\pi}^2}\int dx \,x^2 F' \sin^2F 
=\frac{4}{\pi}\frac{x'^2_0}{e^2f_{\pi}^2}, \;\;
R^2_M [F]= \frac{-1}{2x'^2_0 e^2 f_{\pi}^2}\int dx \,x^4 F' \sin^2F
=\frac{3\pi}{4}\frac{x'^2_0}{e^2f_{\pi}^2}.
\label{93} 
\end{eqnarray}

\begin{table}
\caption{The SU(2)$_f$ Skyrme model isoscalar mean radii $R_{I(M)}$(fm)
and the magnetic moments of the proton and the neutron, in terms of the nucleon Bohr magneton $[\mu_B]$,  
as functions of charge $e$ and for fits (i), (ii) and (iii).}
\begin{center}
\begin{tabular}{|c|cccc|ccccc|ccccc|}
\hline 
${\rm Fits}$ & $ $ & $({\rm i})$ & $ $ & $ $ & $ $ & $ $ & $({\rm ii})$ & $ $ & $ $ & $ $ & $ $ & $({\rm iii})$ & $ $ & $ $  \\
\hline \hline
$ e $ & $R_I$ & $R_M$ & $\mu_p$ & $\mu_n$ & $x'_0$ & $R_I$ & $R_M$ & $\mu_p$ & $\mu_n$ & $x'_0$ & $R_I$ & $R_M$ & $\mu_p$ & $\mu_n$ \\
\hline 
$ 3.0$ & $ 0.773 $ & $ 1.051 $ & $ 6.774 $ & $ -6.563 $ & $ 0.8359 $ & $ 0.667 $ & $ 0.907 $ & $ 5.167 $ & $ -4.956 $ & $ 0.8408 $ & $ 0.671 $
& $ 0.913 $ & $ 5.207 $ & $ -4.996 $ \\
 
$ 3.2$ & $ 0.724 $ & $ 0.985 $ & $ 5.609 $ & $ -5.380 $ & $ 0.8465 $ & $ 0.633 $ & $ 0.862 $ & $ 4.381 $ & $ -4.154 $ & $ 0.8517 $ & $ 0.637 $ 
& $ 0.867 $ & $ 4.418 $ & $ -4.189 $ \\
 
$ 3.4$ & $ 0.682 $ & $ 0.928 $ & $ 4.703 $ & $ -4.459 $ & $ 0.8561 $ & $ 0.603 $ & $ 0.820 $ & $ 3.753 $ & $ -3.509 $ & $ 0.8615 $ & $ 0.607 $ 
& $ 0.825 $ & $ 3.786 $ & $ -3.541 $ \\

$ 3.6$ & $ 0.644 $ & $ 0.876 $ & $ 3.990 $ & $ -3.728 $ & $ 0.8646 $ & $ 0.575 $ & $ 0.782 $ & $ 3.245 $ & $ -2.985 $ & $ 0.8703 $ & $ 0.579 $ 
& $ 0.787 $ & $ 3.275 $ & $ -3.014 $ \\

$ 3.8$ & $0.610 $ & $ 0.830 $ & $ 3.420 $ & $ -3.142 $ & $ 0.8723 $ & $ 0.550 $ & $ 0.748 $ & $ 2.830 $ & $ -2.554 $ & $ 0.8782 $ & $ 0.553 $ 
& $ 0.753 $ & $ 2.857 $ & $ -2.580 $ \\
 
$ 4.0$ & $0.580 $ & $ 0.788 $ & $ 2.960 $ & $ -2.666 $ & $ 0.8792 $ & $ 0.527 $ & $ 0.716 $ & $ 2.488 $ & $ -2.196 $ & $ 0.8854 $ & $ 0.530 $ 
& $ 0.721 $ & $ 2.513 $ & $ -2.219 $ \\

$ 4.1$ & $0.565 $ & $ 0.769 $ & $ 2.763 $ & $ -2.461 $ & $ 0.8824 $ & $ 0.515 $ & $ 0.701 $ & $ 2.340 $ & $ -2.039 $ & $ 0.8887 $ & $ 0.519 $ 
& $ 0.706 $ & $ 2.363 $ & $ -2.061 $ \\
 
$ 4.2$ & $0.552 $ & $ 0.751 $ & $ 2.585 $ & $ -2.275 $ & $ 0.8855 $ & $ 0.505 $ & $ 0.687 $ & $ 2.204 $ & $ -1.896 $ & $ 0.8918 $ & $ 0.508 $ 
& $ 0.692 $ & $ 2.226 $ & $ -1.917 $ \\

$ 4.4$ & $0.527 $ & $ 0.717 $ & $ 2.276 $ & $ -1.950 $ & $ 0.8911 $ & $ 0.485 $ & $ 0.660 $ & $ 1.966 $ & $ -1.642 $ & $ 0.8976 $ & $ 0.488 $ 
& $ 0.664 $ & $ 1.986 $ & $ -1.661 $ \\

$ 4.6$ & $0.504 $ & $ 0.686 $ & $ 2.020 $ & $ -1.679 $ & $ 0.8962 $ & $ 0.466 $ & $ 0.635 $ & $ 1.766 $ & $ -1.426 $ & $ 0.9020 $ & $ 0.470 $ 
& $ 0.639 $ & $ 1.784 $ & $ -1.443 $ \\

$ 4.8$ & $0.483 $ & $ 0.657 $ & $ 1.806 $ & $ -1.449 $ & $ 0.9009 $ & $ 0.449 $ & $ 0.611 $ & $ 1.596 $ & $ -1.241 $ & $ 0.9076 $ & $ 0.453 $ 
& $ 0.616 $ & $ 1.613 $ & $ -1.256 $ \\

$ 5.0$ & $0.464 $ & $ 0.631 $ & $ 1.626 $ & $ -1.254 $ & $ 0.9051 $ & $ 0.433 $ & $ 0.590 $ & $ 1.451 $ & $ -1.081 $ & $ 0.9122 $ & $ 0.437 $ 
& $ 0.594 $ & $ 1.467 $ & $ -1.095 $ \\
\hline \hline 
$ \rm Exp.$ & $ 0.72$ & $ 0.81 $ & 2.79 & $ -1.91 $ & $ - $ & $0.72$ & $ 0.81 $ & 2.79 & $ -1.91 $ & $ - $ & $0.72$ 
& $ 0.81 $ & 2.79 & $ -1.91 $ \\
\hline \hline 
\end{tabular}
\label{t:tab5}
\end{center}
\end{table}
\renewcommand{\arraystretch}{1}
To obtain the {\bf 8}, {\bf 10}, $\overline{\bf 10}$ and ${\bf 27}_{3/2}$ 
mass spectrums we use the mass formula (\ref{61}) and find 
\begin{eqnarray}
M^{\bf 8}_B[F] 
&=& {\cal M}_{\bf 8} - \frac{1}{2}\gamma(x'_0) \;\delta^{\bf 8}_B, \;\;
M^{\bf 10}_B[F] 
= {\cal M}_{\bf 8} +\frac{3}{2\lambda_c (x_0^{\prime})}- \frac{1}{2}\gamma(x'_0)\;\delta^{\bf 10}_B, 
\label{95}\\
M^{\overline{\bf 10}}_B[F] 
&=&{\cal M}_{\bf 8} +\frac{3}{2\lambda_s (x_0^{\prime})}-\frac{1}{2}\gamma(x'_0)\;\delta^{\overline{\bf 10}}_B,\;\;
{\rm M}^{\bf 27}_B[F]= {\cal M}^{\bf 8} 
+\frac{3}{2\lambda_c (x_0^{\prime})}+\frac{1}{\lambda_s (x_0^{\prime})}-\frac{1}{2}\gamma(x_0^{\prime})\;\delta^{\bf 27}_{B},
\label{97}
\end{eqnarray}
where the experimental octet mean mass 
${\cal M}^{\bf 8}=\frac{1}{8}\sum_{B=1}^{\bf 8} M^{\bf 8}_B=1151$ MeV
was used instead of 
\begin{eqnarray}
{\cal M}^{\bf 8}={\cal M}_{\rm csol}(x_0^{\prime})+\frac{3}{2\lambda_c (x_0^{\prime})}. 
\label{98}
\end{eqnarray}
The reason is simply because nowadays everybody agrees that the SU(3)$_f$ extended Skyrme model 
classical soliton mass ${\cal M}_{\rm csol}$ receives to large value producing unrealistic baryonic mass spectrum. 
From measurements we also know
${\cal M}^{\bf 10}=\frac{1}{10}\sum^{10}_{B=1} M^{\bf 10}_B =1382$ MeV \cite{rpp}.  
The splitting constants $\delta^{\bf 8}_B$, $\delta^{\bf 10}_B$ and $\delta^{\overline{\bf 10}}_B$  
are given in Eqs. (5) to (18) of Ref. \cite{Ell}, while $\delta^{\bf 27}_B$ 
could be found in Table 1. of Ref. \cite{WM}. 

Equations (\ref{95},\ref{97}) assume equal spacing between multiplets.
From the existing experiments \cite{penta1}--\cite{penta6}
($M_{\Theta^+}=1540$ MeV and $M_{\Xi^{--}_{3/2}}=1861$ MeV) we estimate that spacing to be 
$\overline{\delta}=(1861-1540)/3=107$ MeV.
Next we estimate masses of antidecuplets $M_{N^{*}}=1647$ MeV, $M_{\Sigma_{\overline{10}}} =1754$ MeV and the  
${\overline{\bf 10}}$ mean mass 
${\cal M}_{\overline{\bf 10}}=\frac{1}{10}\sum^{10}_{B=1}M^{\overline{\bf 10}}_B =1754$ MeV,
and using them bonafide as an ``experimental'' values further on. 
From the expressions (\ref{95},\ref{97}) it is clear that in the SU(3)$_f$ symmetric
case and in the chiral limit the {\bf 8}, {\bf 10}, $\overline{\bf 10}$ and ${\bf 27}_{3/2}$, 
absolute masses of each member of the multiplet become equal for each fixed $e$: 
$\rm N=\Lambda=\Sigma=\Xi \equiv M^{\bf 8}_B$,
$\Delta=\Sigma^*=\Xi^*=\Omega \equiv M^{\bf 10}_B$, 
$\Theta^+=\rm N^*=\Sigma_{\overline{10}}=\Xi^{--}_{3/2}\equiv M^{\overline{\bf 10}}_B$ and
$\Theta_1={\rm N}^*_{\frac{3}{2}}= \Sigma_2={\rm N}^*_{\frac{1}{2}}= 
\Sigma_1= \Lambda^*=\Xi^*_{\frac{3}{2}}=\Xi^*_{\frac{1}{2}}=\Omega_1\equiv M^{\bf 27}_B$.
For example, with ${\cal M}_{\bf 8}=1151$ MeV as an input and for $e=4.7$ we would have 
$\left[M^{\bf 8}_B,M^{\bf 10}_B,M^{\overline{\bf 10}}_B,M^{\bf 27}_B\right]=[1151,1422,1865,1898]$ MeV.
Numerics for the {\bf 8}, {\bf 10}, $\overline{\bf 10}$ and ${\bf 27}_{3/2}$ mass spectrums 
in the SU(3)$_f$ Skyrme model as functions of charge $4 \leq e \leq 5$ and for fits (ii) and (iii)
are given in Tables \ref{t:tab6} and \ref{t:tab7}, respectively. 
The Skyrme charge $e$ and the SB effects dependences of the mass spectrums are very transparently
presented in Figures \ref{fig5}, \ref{fig6} and \ref{fig7}, respectively.
Note that in the computations of the mean masses ${\cal M}^{\bf 8}$,...,
the sum of $D_{88}$ diagonal elements over all components of irreducible representations
cancels out because of the properties of the SU(3) Clebsch-Gordan coefficients \cite{desw,fin}.
In the above notation we are following Fig. 4 from Ref.\cite{Ell} as close as possible. However,
the $\overline{\bf 10}$-plet members $\Theta$, $\Sigma$ and $\Xi_{3/2}$ 
we mark as $\Theta^+$, $\Sigma_{\overline{\bf 10}}$ and $\Xi^{--}_{3/2}$, respectively.
The $\Xi$ isoquartet and isodoublet from the ${\bf 27}_{3/2}$ we mark as
$\Xi^*_{\frac{3}{2}}$ and $\Xi^*_{\frac{1}{2}}$, 
to distinguish them from the $\Xi$ isoquartet and isodoublet 
from the $\overline{\bf 10}$. We also mark the ${\bf 27}_{3/2}$-plet isosinglet as $\Lambda^*$. 

\renewcommand{\arraystretch}{1.2}
\begin{table}
\caption{The {\bf 8} and {\bf 10} mass spectrums (MeV) 
as functions of charge $e$ and for fits (ii)and (iii).
The experimental numbers for {\bf 8} and {\bf 10} masses were used from \cite{rpp}}.
\begin{center}
\begin{tabular}{|c|cccccccc|cccccccc|}
\hline
${\rm Fits}$ & $$ & $$ & $$ & $({\rm ii})$ & $$ & $$ & $$ & $$ & $$ & $$ & $$ &$({\rm iii})$ & $$ & $$ & $$ & $$ \\
\hline \hline
$e$ & $\rm N$ & $\Lambda$ & $\Sigma$ & $\Xi$ & $\Delta$ & $\Sigma^*$ & $\Xi^*$ & $\Omega$
& $\rm N$ & $\Lambda$ & $\Sigma$ & $\Xi$ & $\Delta$ & $\Sigma^*$ & $\Xi^*$ & $\Omega$\\
\hline 
$4.0$ & $ 1007 $ & $ 1103$ & $ 1199 $ & $ 1247$ & $ 1291 $ &  $ 1351 $ & $ 1411 $ & $ 1472 $
& $ 907 $ & $ 1070$ & $ 1232 $ & $ 1313$ & $ 1248 $ &  $ 1350 $ & $ 1451 $ & $ 1552 $\\

$4.1$ & $ 1016 $ & $ 1106$ & $ 1196 $ & $ 1241$ & $ 1309 $ & $ 1365 $ &  $ 1422 $ & $ 1478 $
& $ 921 $ & $ 1074$ & $ 1228 $ & $ 1304$ & $ 1268 $ & $ 1363 $ &  $ 1459 $ & $ 1555 $\\

$ 4.2$ & $ 1024 $ & $ 1109$ & $ 1193 $ & $ 1236$ & $ 1327 $ & $ 1380 $ &  $ 1433 $ & $ 1486 $
& $ 934 $ & $ 1079$ & $ 1223 $ & $ 1295$ & $ 1287 $ & $ 1378 $ &  $ 1468 $ & $ 1558 $\\

$4.4$ & $ 1038 $ & $ 1113$ & $ 1189 $ & $ 1226$ & $ 1364 $ & $ 1411 $ &  $ 1458 $ & $ 1505 $
& $ 957 $ & $ 1086$ & $ 1216 $ & $ 1280$ & $ 1328 $ & $ 1408 $ &  $ 1489 $ & $ 1570 $\\
 
$4.6$ & $ 1051 $ & $ 1118$ & $ 1184 $ & $ 1218$ & $ 1403 $ & $ 1445 $ &  $ 1487 $ & $ 1529 $
& $ 977 $ & $ 1093$ & $ 1209 $ & $ 1267$ & $ 1370 $ & $ 1442 $ &  $ 1514 $ & $ 1587 $\\

$4.8$ & $ 1061 $ & $ 1121$ & $ 1181 $ & $ 1211$ & $ 1445 $ & $ 1482 $ &  $ 1519 $ & $ 1557 $
& $ 994 $ & $ 1099$ & $ 1203 $ & $ 1255$ & $ 1413 $ & $ 1478 $ &  $ 1544 $ & $ 1609 $\\

$5.0$ & $ 1071 $ & $ 1124$ & $ 1178 $ & $ 1205$ & $ 1488 $ & $ 1522 $ &  $ 1555 $ & $ 1589 $
& $ 1009 $ & $ 1104$ & $ 1198 $ & $ 1246$ & $ 1458 $ & $ 1518 $ &  $ 1577 $ & $ 1636 $\\
\hline \hline 
$\rm Exp.$ & $ 939$ & $ 1116$ & $ 1193 $ & $ 1318$ & $ 1232 $ & $ 1385 $ & $1530$ & $ 1672$
& $ 939$ & $ 1116$ & $ 1193 $ & $ 1318$ & $ 1232 $ & $ 1385 $ & $1530$ & $ 1672$\\
\hline\hline
\end{tabular}
\label{t:tab6}
\end{center}
\end{table}

\renewcommand{\arraystretch}{1.4}
\begin{table}
\caption{The $\overline{\bf 10}$ and {\bf 27}$_{3/2}$ mass spectrums (MeV) 
as functions of Skyrme charge $e$ and fits (ii), (iii).
The $\Theta^+$ and $\Xi^{--}_{3/2}$ experimental masses can be find in \cite{penta1}--\cite{penta6}.}
\begin{center}
\begin{tabular}{|c|c|cccc|ccccccccc|}
\hline
${\rm Fits}$ & $ e $ & $\Theta^+$ & $\rm N^*$ & $\Sigma_{\overline{10}}$ &
$\Xi^{--}_{3/2}$ & $\Theta_1$ & $\rm N^*_{\frac{3}{2}}$ & $\Sigma_2$ & $\rm N^*_{\frac{1}{2}}$ & $\Sigma_1$ & 
$\Lambda^*$ & $\Xi^*_{\frac{3}{2}}$ & $\Xi^*_{\frac{1}{2}}$ & $\Omega_1$ \\
\hline \hline
$  $ & $ 4.0 $
& $ 1576 $ & $ 1636 $ & $ 1696 $ & $ 1756 $ 
& $ 1646$ & $ 1659 $ & $ 1671$ & $ 1697 $ &  $ 1723 $ & $ 1749 $ & $ 1749 $ & $ 1788 $ & $ 1826 $ \\

$  $ & $ 4.1 $
& $ 1620 $ & $ 1677 $ & $ 1733 $ & $ 1789 $ 
& $ 1689$ & $ 1701 $ & $ 1713$ & $ 1737 $ &  $ 1761 $ & $ 1786 $ & $ 1786 $ & $ 1822 $ & $ 1858 $ \\

$(\rm{ii})$ & $ 4.2 $
& $ 1666 $ & $ 1719 $ & $ 1772 $ & $ 1825 $ 
& $ 1734$ & $ 1745 $ & $ 1756$ & $ 1779 $ &  $ 1802 $ & $ 1824 $ & $ 1824 $ & $ 1858 $ & $ 1892 $ \\

$ $ & $ 4.4 $
& $ 1761 $ & $ 1808 $ & $ 1855 $ & $ 1902 $ 
& $ 1827$ & $ 1837 $ & $ 1847$ & $ 1867 $ &  $ 1887 $ & $ 1908 $ & $ 1908 $ & $ 1938 $ & $ 1968 $ \\
 
$  $ & $ 4.6 $
& $ 1862 $ & $ 1904 $ & $ 1946 $ & $ 1988 $ 
& $ 1927$ & $ 1936 $ & $ 1945$ & $ 1963 $ & $ 1981 $ &  $ 1999 $ & $ 1999 $ & $2026 $ & $ 2053 $  \\
 
$  $ & $ 4.8 $
& $ 1969 $ & $ 2007 $ & $ 2044 $ & $ 2082 $ 
& $ 2035$ & $ 2043 $ & $ 2051$ & $ 2067 $ & $ 2083 $ &  $ 2099 $ & $ 2099 $ & $ 2123 $ & $ 2147 $ \\

$  $ & $ 5.0 $
& $ 2083 $ & $ 2116 $ & $ 2150 $ & $ 2183 $ 
& $ 2149$ & $ 2157 $ & $ 2164$ & $ 2178 $ &  $ 2193 $ & $ 2207 $ & $ 2207 $ & $ 2229 $ & $ 2250 $ \\
\hline
$  $ & $ 4.0 $
& $ 1364 $ & $ 1466 $ & $ 1567 $ & $ 1668 $ 
& $ 1511$ & $ 1533 $ & $ 1554$ & $ 1598 $ &  $ 1642 $ & $ 1685 $ & $ 1685 $ & $ 1750 $ & $ 1816 $ \\

$  $ & $ 4.1 $
& $ 1404 $ & $ 1500 $ & $ 1595 $ & $ 1691 $ 
& $ 1550$ & $ 1571 $ & $ 1591$ & $ 1532 $ &  $ 1673 $ & $ 1714 $ & $ 1714 $ & $ 1776 $ & $ 1837 $ \\

$(\rm{iii})$ & $ 4.2 $
& $ 1444 $ & $ 1535 $ & $ 1625 $ & $ 1715 $ 
& $ 1590$ & $ 1610 $ & $ 1629$ & $ 1668 $ & $ 1706 $ &  $ 1745 $ & $ 1745 $ & $ 1803 $ & $ 1861 $ \\

$ $ & $ 4.4 $
& $ 1526 $ & $ 1607 $ & $ 1687 $ & $ 1768 $ 
& $ 1674$ & $ 1691 $ & $ 1708$ & $ 1743 $ & $ 1778 $ &  $ 1812 $ & $ 1812 $ & $ 1864 $ & $ 1916 $ \\

$  $ & $ 4.6 $
& $ 1611 $ & $ 1683 $ & $ 1755 $ & $ 1828 $ 
& $ 1762$ & $ 1778 $ & $ 1793$ & $ 1824 $ & $ 1855 $ &  $ 1886 $ & $ 1886 $ & $ 1933 $ & $ 1979 $ \\
 
$  $ & $ 4.8 $
& $ 1699 $ & $ 1764 $ & $ 1829 $ & $ 1894 $ 
& $ 1856$ & $ 1870 $ & $ 1884$ & $ 1912 $ & $ 1940 $ &  $ 1968 $ & $ 1968 $ & $ 2010 $ & $ 2052 $ \\

$  $ & $ 5.0 $
& $ 1791 $ & $ 1850 $ & $ 1909 $ & $ 1968 $ 
& $ 1955$ & $ 1969 $ & $ 1981$ & $ 2006 $ &  $ 2031 $ & $ 2057 $ & $ 2057 $ & $ 2094 $ & $ 2133 $ \\
\hline\hline
$\rm Exp.$ & $ $ & 
$1540$ & $-$ & $-$ & $1861$ & $-$ & $-$ & $-$ & $-$ & $-$ & $-$ & $-$ & $-$ & $-$ \\
\hline \hline
\end{tabular}
\label{t:tab7}
\end{center}
\end{table}
\renewcommand{\arraystretch}{1}
The predictions for higher SU(3)$_f$ representations mass splittings are in order. 
The mass splittings $\Delta_i[F],\, i=1,...,4$, for the multiplets 
${\bf 8}_{J=1/2}$, ${\bf 10}_{J=3/2}$, $\overline{\bf 10}_{J=1/2}$, 
${\bf 27}_{J=3/2}$ and ${\bf 35}_{J=5/2}$
expressed by the following simple relations:
\begin{eqnarray}
{\Delta}_1 [F] &=& {\cal M}^{\bf 10} - {\cal M}^{\bf 8}= 
\frac{3}{2\lambda_c (x_0^{\prime})}\equiv {\Delta}_1,\;\;\;\,
\hspace{2mm}{\Delta}_2 [F] = {\cal M}^{\overline{\bf 10}} - {\cal M}^{\bf 8}=
\frac{3}{2\lambda_s (x_0^{\prime})} \equiv {\Delta}_2, 
\nonumber\\
{\Delta}_3 [F] &=&{\cal M}^{\bf 27}_{3/2} - {\cal M}^{\overline{\bf 10}}= 
{\Delta}_1-\frac{1}{3}{\Delta}_2 ,\;\,
\hspace{8mm}{\Delta}_4[F]={\cal M}^{\bf 35}_{5/2}-{\cal M}^{\bf 27}_{3/2}=\frac{5}{3}{\Delta}_1-\frac{1}{3}{\Delta}_2, 
\label{99} 
\end{eqnarray}
are evaluated as functions of Skyrme charge $e$ and 
for three sets of parameters, (i), (ii) and (iii), and presented in Table \ref{t:tab8}.
In the chiral limit and for $e=4.7$ we would have:
\begin{eqnarray}
{\Delta}_3
&=&\left[\frac{52e^3f_{\pi}}{285\sqrt{30}\pi^2}\right]_{e=4.7} = 32.6\; \rm MeV 
\label{100}\\
{\Delta}_4
&=&\left[\frac{68e^3f_{\pi}}{57\sqrt{30}\pi^2}\right]_{e=4.7} = 213.1\; \rm MeV, \;{\rm and}\;
\left[{\cal M}^{\bf 27}_{3/2}\right]_{e=4.7}=1898\;\rm MeV.
\nonumber
\end{eqnarray}
Cases (ii) and (iii) are graphically presented in Figures \ref{fig8} and \ref{fig9}, respectively.

All other mass splittings $\Delta_i(e,{\hat x},\beta^{\prime},\delta^{\prime}),\, i=5,...,12$, 
for all combinations of the multiplets, including members of penta-quark family 
($\overline{\bf 10},{\bf 27},{\bf 35}$) 
and lowest members of the septu-quark families $\overline{\bf 35}$ and $\bf 64$ are expressed in terms of 
mass splittings $\Delta_1$ and $\Delta_2$:
\begin{eqnarray}
{\Delta}_5 [F] &=&{\cal M}^{\bf 35}_{5/2} - {\cal M}^{\overline{\bf 10}}= 
\frac{8}{3}{\Delta}_1-\frac{2}{3}{\Delta}_2,\;
\hspace{6mm}{\Delta}_6[F] ={\cal M}^{\overline{\bf 10}} - {\cal M}^{\bf 10}= -{\Delta}_1+{\Delta}_2, 
\label{101} \\
{\Delta}_7 [F] &=&{\cal M}^{\bf 27}_{1/2} - {\cal M}^{\bf 8}= \frac{5}{3}{\Delta}_2 ,\;\,
\hspace{19mm}{\Delta}_8[F]={\cal M}^{\bf 27}_{3/2}-{\cal M}^{\bf 10}=\frac{2}{3}{\Delta}_2, 
\nonumber \\
{\Delta}_9 [F] &=&{\cal M}^{\overline{\bf 35}}_{3/2} - {\cal M}^{\bf 35}_{5/2}= \frac{5}{3}{\Delta}_6,\;
\hspace{17mm}{\Delta}_{10} [F] ={\cal M}^{\overline{\bf 35}}_{3/2} - {\cal M}^{\bf 27}_{3/2}= \frac{4}{3}{\Delta}_2. 
\nonumber \\
{\Delta}_{11} [F] &=&{\cal M}^{\overline{\bf 35}}_{3/2} - {\cal M}^{\bf 10}= \frac{5}{2}{\Delta}_2,\;
\hspace{18mm}{\Delta}_{12} [F] ={\cal M}^{\bf 64}_{3/2} - {\cal M}^{\bf 27}_{3/2}= \frac{7}{3}{\Delta}_2. 
\nonumber 
\end{eqnarray}

The mass splittings between minimal and non-minimal multiplets depend on $\lambda_s$ and on linear combinations of
$\lambda_c$ and $\lambda_s$, while mass splittings between minimal multiplets 
({\bf 8} and {\bf 10}) depend on $\lambda_c$ only.
  
Combining experiments ($M_{\Theta^+}=1540$ MeV and $M_{\Xi^{--}_{3/2}}=1861$ MeV) and 
earlier estimates of the ``experimental'' antidecuplet masses $M_{N^*}$=1647 MeV, 
$M_{\Sigma_{\overline{10}}} =1754$ MeV
and the ${\overline{\bf 10}}$ mean mass ${\cal M}^{\overline{\bf 10}} =1754$ MeV
we obtain the antidecuplet--octet mass splitting
${\Delta}_2^{\rm exp} ={\cal M}^{\overline{\bf 10}}-{\cal M}^{\bf 8}=603$ MeV, the value which we are using
bonafide as an ``experimental'' number further on.
However, the decuplet--octet mass splitting ${\Delta}_1^{\rm exp} =231$ MeV represent the true experimental value.

Owing to the cancellation between $\Delta_1$ and $\Delta_2$ the mass splittings $\Delta_3$ and $\Delta_4$
represent the smallest among all of the splittings (\ref{99},\ref{101}) 
between the SU(3)$_f$ multiplets 
${\bf 8}$, ${\bf 10}$, $\overline{\bf 10}$, ${\bf 27}$, ${\bf 35}$, $\overline{\bf 35}$
and $\bf 64$. 
\renewcommand{\arraystretch}{1.4}
\begin{table}
\caption{The mass splittings $\Delta_i(e,{\hat x},\beta^{\prime},\delta^{\prime}),\;i=1,...,4$, 
(MeV) as functions of charge $e$ and for fits (i), (ii) and (iii).
Experimental number for $\Delta_1$ was used from \cite{rpp}.}
\begin{center}
\begin{tabular}{|c|cccc|cccc|cccc|}
\hline
$ {\rm Fits} $ & $ $ & $({\rm i})$& $ $ & $ $ & $ $ & $({\rm ii})$& $ $ & $ $ & $ $ & $({\rm iii})$& $ $ & $ $ \\
\hline \hline 
$ e $ & $\Delta_1$ & $\Delta_2$ & $\Delta_3$& $\Delta_4$ & $\Delta_1$ & $\Delta_2$ & $\Delta_3$& $\Delta_4$ & 
$\Delta_1$ & $\Delta_2$ & $\Delta_3$& $\Delta_4$ \\
\hline 
$ 4.0$ & $ 166.9 $ & $ 440.4$& $20.1$& $131.4$ & $ 200.5 $ & $ 544.7 $& $18.9$& $152.5$
 &  $ 198.4 $ & $ 416.3 $& $59.7$& $192.0$ \\

$ 4.1$ & $ 179.7 $ & $ 474.3$& $21.6$& $141.5$ & $ 214.4 $ & $ 582.0 $& $20.4$& $163.3$
 &  $ 212.2 $ & $ 444.4 $& $64.1$& $205.6$ \\
 
$ 4.2$ & $ 193.1 $ & $ 509.8$& $23.2$& $152.1$ & $ 229.0 $ & $ 621.0 $& $22.0$& $174.7$
 &  $ 226.7 $ & $ 473.8 $& $68.7$& $219.8$ \\
 
$ 4.4$ & $ 222.1 $ & $ 586.2$& $26.7$& $174.8$ & $ 260.2 $ & $ 704.3 $& $25.5$& $198.9$
 &  $ 257.4 $ & $ 536.4 $& $78.6$& $250.3$ \\

$ 4.6$ & $ 253.8 $ & $ 669.8$& $30.5$& $199.8$ & $ 294.2 $ & $ 794.8 $& $29.3$& $225.4$
 &  $ 290.1 $ & $ 604.5 $& $89.5$& $283.4$ \\

$ 4.8$ & $ 288.4 $ & $ 761.0$& $34.7$& $227.0$ & $ 331.0 $ & $ 892.9 $& $33.4$& $254.1$
 &  $ 327.3 $ & $ 678.2 $& $101.2$& $319.4$ \\

$ 5.0$ & $ 326.0 $ & $ 860.2$& $39.3$& $256.5$ & $ 370.8 $ & $ 998.9 $& $37.9$& $285.1$
 &  $ 366.6 $ & $ 757.8 $& $114.0$& $358.3$ \\
\hline \hline 
$\rm Exp.$ & $231$ & $-$& $-$& $-$ & $231$ & $-$& $-$& $-$ & $231$ & $-$& $-$& $-$ \\
\hline \hline 
\end{tabular}
\label{t:tab8}
\end{center}
\end{table}

\renewcommand{\arraystretch}{1.4}
\begin{table}
\caption{The {\bf 27}$_{3/2}-\overline{\bf 10}$ mass splittings (MeV) 
as functions of Skyrme charge $e$ and for fits (ii), (iii).}
\begin{center}
\begin{tabular}{|c|cccccccc|cccccccc|}
\hline
${\rm Fits}$ & $$ & $$ & $$ & $({\rm ii})$ & $$ & $$ & $$ & $$ & $$ & $$ & $$ &$({\rm iii})$ & $$ & $$ & $$ & $$ \\
\hline \hline
$ e $ 
& $\delta_1$ & $\delta_2$ & $\delta_3$ & $\delta_4$ & $\delta_5$ & $\delta_6$ & $\delta_7$ & $\delta_8$ 
& $\delta_1$ & $\delta_2$ & $\delta_3$ & $\delta_4$ & $\delta_5$ & $\delta_6$ & $\delta_7$ & $\delta_8$ \\
\hline 
$ 4.0 $ & $ 70$ & $ 23 $ & $ 62$ & $ -24 $ &  $ 27 $ & $ 53 $ & $ -7 $ & $ 32 $ 
& $ 147$ & $ 67 $ & $ 132$ & $ -13 $ &  $ 74 $ & $ 118 $ & $ 16 $ & $ 81 $\\

$ 4.1 $ & $ 69$ & $ 24 $ & $ 61$ & $ -20 $ &  $ 28 $ & $ 53 $ & $ -4 $ & $ 32 $ 
& $ 146$ & $ 71 $ & $ 132$ & $ -4 $ &  $ 78 $ & $ 119 $ & $ 23 $ & $ 85 $\\

$ 4.2 $ & $ 67$ & $ 26 $ & $ 60$ & $ -16 $ &  $ 30 $ & $ 53 $ & $ -1 $ & $ 33 $ 
& $ 146$ & $ 75 $ & $ 133$ & $ 4 $ & $ 82 $ &  $ 121 $ & $ 30 $ & $ 88 $\\

$ 4.4 $ & $66$ & $ 29 $ & $ 59$ & $ -8 $ &  $ 32 $ & $ 52 $ & $ 5 $ & $ 36 $ 
& $ 148$ & $ 84 $ & $136$ & $ 21 $ & $ 90 $ &  $ 125 $ & $ 44 $ & $ 96$\\
 
$ 4.6 $ & $ 65$ & $ 32 $ & $ 59$ & $ -1$ & $ 35 $ &  $ 53 $ & $ 11 $ & $ 38 $ 
& $ 151$ & $ 95 $ & $ 141$ & $ 38 $ & $ 100 $ &  $ 131 $ & $ 58 $ & $ 105 $\\
 
$ 4.8 $ & $ 65$ & $ 36 $ & $ 60$ & $ 7 $ & $ 39 $ &  $ 55 $ & $ 17 $ & $ 41 $ 
& $ 157$ & $ 106 $ & $ 148$ & $ 55 $ & $ 111 $ &  $ 139 $ & $ 73 $ & $ 115 $\\

$ 5.0 $ & $ 67$ & $ 40 $ & $ 62$ & $ 14 $ &  $ 43 $ & $ 57 $ & $ 24 $ & $ 45 $ 
& $ 165$ & $ 118 $ & $ 156$ & $ 72 $ &  $ 122 $ & $ 148 $ & $ 89 $ & $ 127 $\\
\hline \hline
\end{tabular}
\label{t:tab9}
\end{center}
\end{table}
\renewcommand{\arraystretch}{1}
Next we present the splittings between the same quark content baryons of ${\bf 27}_{3/2}$ and 
$\overline{\bf 10}$ representations: 
\begin{eqnarray}
\delta_1
&=&M^{\bf 27}_{3/2}(\Theta_1)-M^{\overline{\bf 10}}(\Theta^+)
={\Delta}_3+\frac{3}{56}\gamma,\;\;\;
\hspace{2mm}\delta_2
=M^{\bf 27}_{3/2}(N^*_{\frac{3}{2}})-M^{\overline{\bf 10}}(N^*)
={\Delta}_3+\frac{1}{224}\gamma,
\label{102}\\
\delta_3
&=&M^{\bf 27}_{3/2}(N^*_{\frac{1}{2}})-M^{\overline{\bf 10}}(N^*)
={\Delta}_3+\frac{5}{112}\gamma,\;\;\;
\delta_4
=M^{\bf 27}_{3/2}(\Sigma_2)-M^{\overline{\bf 10}}(\Sigma_{\overline{10}})
={\Delta}_3-\frac{5}{112}\gamma,
\nonumber\\
\delta_5
&=&M^{\bf 27}_{3/2}(\Sigma_1)-M^{\overline{\bf 10}}(\Sigma_{\overline{10}})
={\Delta}_3+\frac{1}{112}\gamma,\;\;\;
\hspace{.2mm}\delta_6
=M^{\bf 27}_{3/2}(\Lambda^*)-M^{\overline{\bf 10}}(\Sigma_{\overline{10}})
={\Delta}_3+\frac{1}{28}\gamma,
\nonumber\\
\delta_7
&=&M^{\bf 27}_{3/2}(\Xi^*_{\frac{3}{2}})-M^{\overline{\bf 10}}(\Xi^{--}_{3/2})
={\Delta}_3-\frac{3}{112}\gamma,\;\;\;
\hspace{1.4mm}\delta_8
=M^{\bf 27}_{3/2}(\Xi^*_{\frac{1}{2}})-M^{\overline{\bf 10}}(\Xi^{--}_{3/2})
={\Delta}_3+\frac{3}{224}\gamma.
\nonumber
\end{eqnarray}
Due to the absence of anomalous moments of inertia \cite{HW}, 
$M^{\bf 27}_{3/2}(\Theta_1)-M^{\overline{\bf 10}}(\Theta^+)
=M^{\bf 27}_{3/2}(\Omega_1)-M^{\overline{\bf 10}}(\Xi^{--}_{3/2})$
and $M^{\bf 27}_{3/2}(\Xi^*_{\frac{3}{2}})-M^{\overline{\bf 10}}(\Xi^{--}_{3/2})
=M^{\bf 27}_{3/2}(\Lambda^*)-M^{\overline{\bf 10}}(\Xi^{--}_{3/2})$.
Mass differences $\delta_{1,...,8}$,
as functions of the Skyrme charge $e$ and fits (ii, iii) are given in Table \ref{t:tab9}
and are graphically presented in Figures \ref{fig10}, \ref{fig11}, respectively.
In the chiral limit, fit (i), $\delta_{1,...,8}=\Delta_3$.

\section{Discussions and conclusions}

For $f_{\pi}^{\rm exp}=93$ MeV and for the particular value of $e=4.1$, which is favored in the 
SU(3)$_f$ extension of the Skyrme model with SB
terms included \cite{wei}, in the chiral limit of the SU(2)$_f$ Skyrme model, 
the nucleon axial-vector coupling (\ref{44}) and proton magnetic moment (\ref{47}) are
\begin{eqnarray}
g_A = 1.25,\;\;\Delta_1=179.7\;\rm MeV,\;\;R_I = 0.57 \;\rm fm, \; \; 
R_M = 0.77 \;\rm fm, \;\;
\mu_p = 2.76\,\mu_B , \; \; \mu_n = -2.46\,\mu_B,
\label{103}
\end{eqnarray}
in excellent agreement with measurements. 
The other static properties are more or less close to the experimental values.
However, the extension of SU(2)$_f\, \to {\rm SU(3)}_f$ introduces nontrivial Clebsch-Gordan coefficients
which erase the nice agreement with experiment of the $g_A$ and $\mu_p$ indicating that under the SU(3)$_f$
dynamics exist other effects associated with possible admixture in
total baryon wave function producing additional contributions.  

\begin{figure}[h]
\centerline{\includegraphics{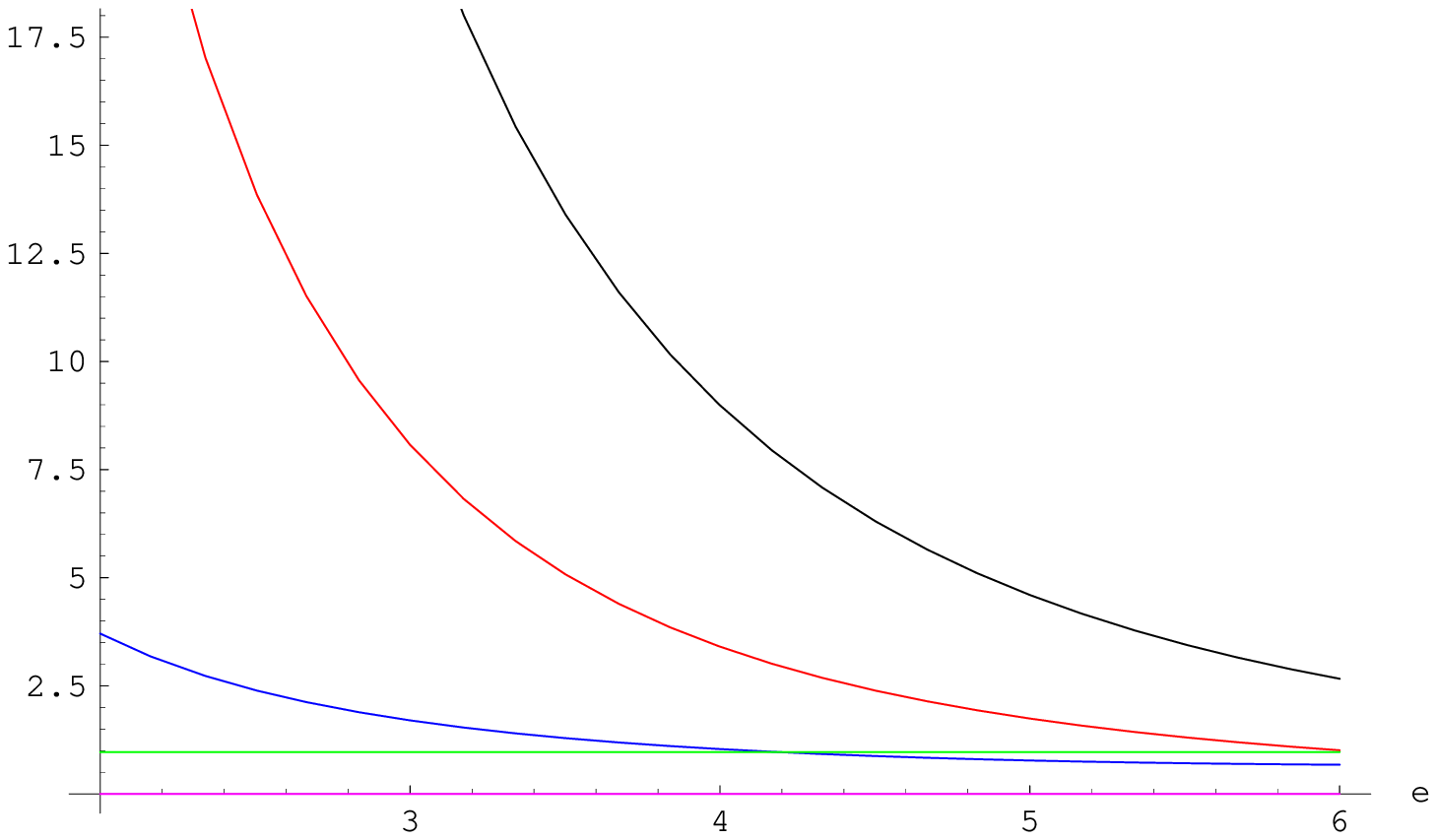}}
\caption{Fit (i): The SB quantity $\gamma$ (pink) in GeV, dimensionles size of soliton $x'_0$ (green), 
nucleon axial-vector constant $g_A$ (blue),
moments of inertia $\lambda_s$ (red) in GeV$^{-1}$ and $\lambda_c$ (black) in GeV$^{-1}$.}
\label{fig2}
\end{figure}

\begin{figure}[h]
\centerline{\includegraphics{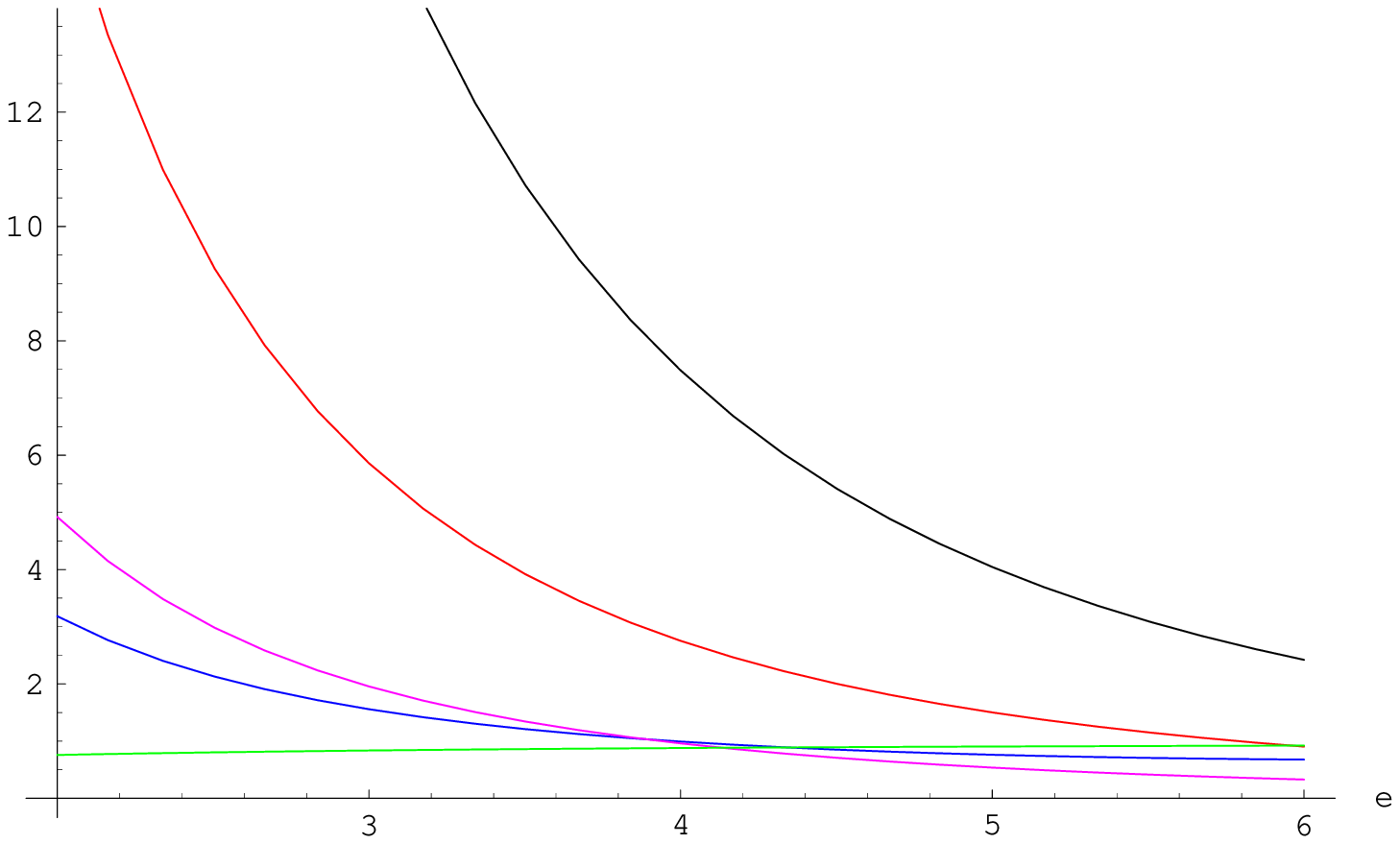}}
\caption{Fit (ii): Dimensionles size of soliton $x'_0$ (green), nucleon axial-vector constant $g_A$ (blue),
the SB quantity $\gamma$ (pink) in GeV, 
moments of inertia $\lambda_s$ (red) in GeV$^{-1}$ and $\lambda_c$ (black) in GeV$^{-1}$.}
\label{fig3}
\end{figure}

\begin{figure}[h]
\centerline{\includegraphics{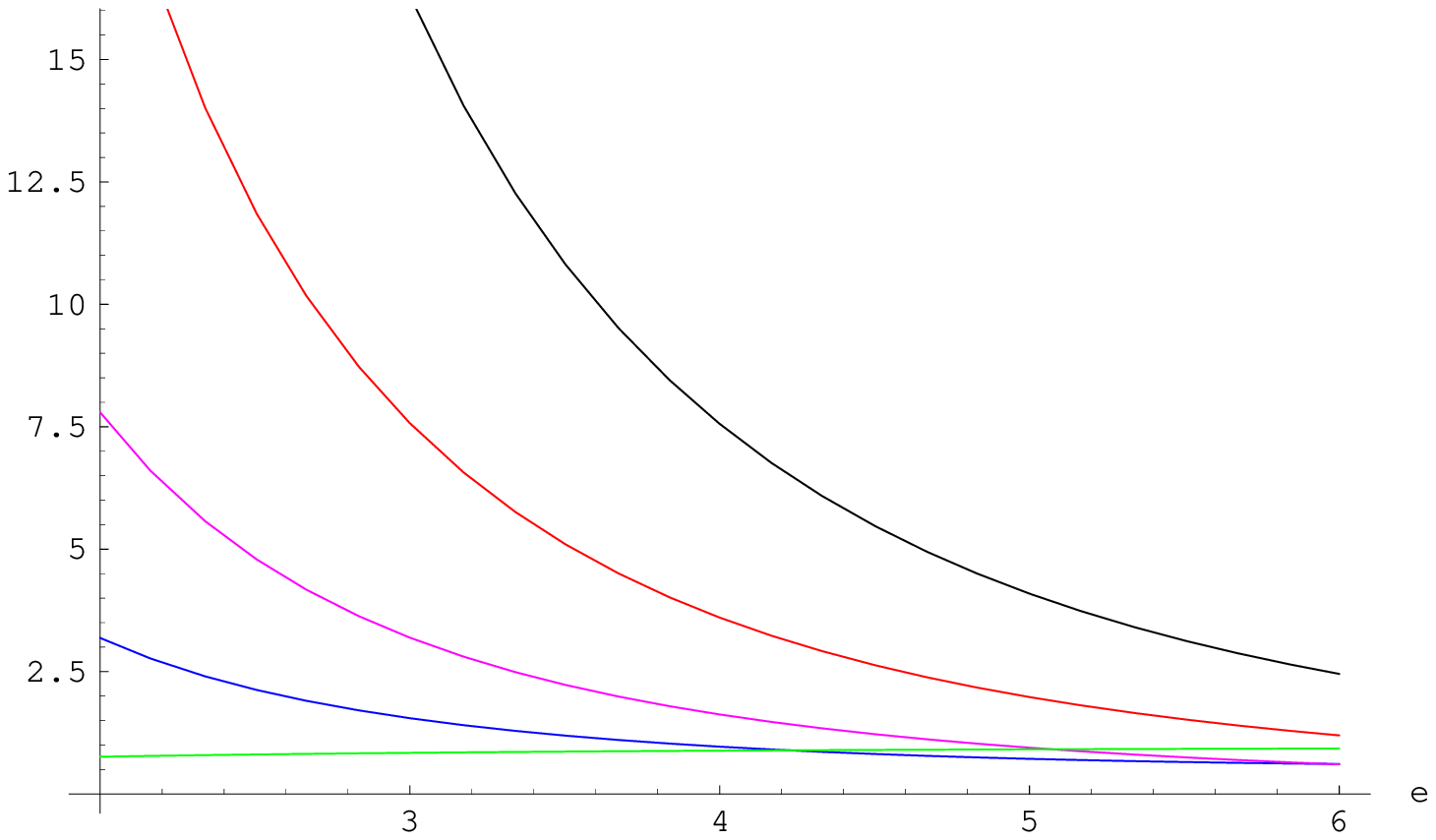}}
\caption{Fit (iii): Dimensionles size of soliton $x'_0$ (green), nucleon axial-vector constant $g_A$ (blue),
the SB quantity $\gamma$ (pink) in GeV, 
moments of inertia $\lambda_s$ (red) in GeV$^{-1}$ and $\lambda_c$ (black) in GeV$^{-1}$.}
\label{fig4}
\end{figure}

\begin{figure}[h]
\centerline{\includegraphics{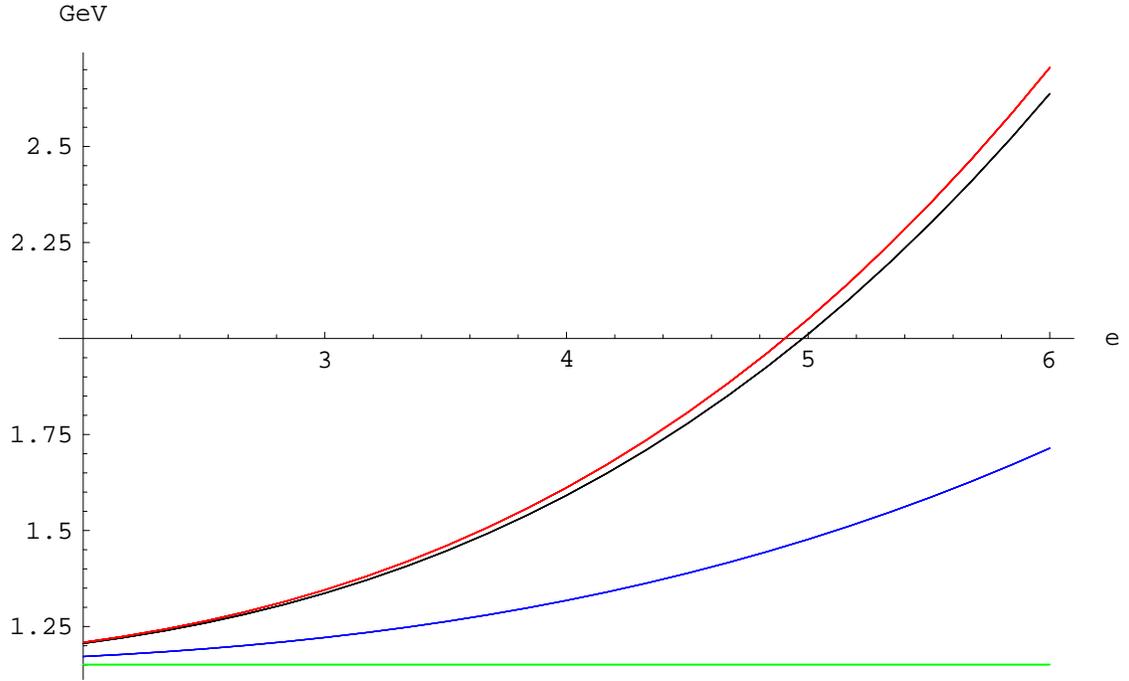}}
\caption{Fit (i): Mass spectrums for {\bf 8} (green), {\bf 10} (blue), 
$\overline{\bf 10}$ (black) and ${\bf 27}_{3/2}$ (red).}
\label{fig5}
\end{figure}

\begin{figure}[h]
\centerline{\includegraphics{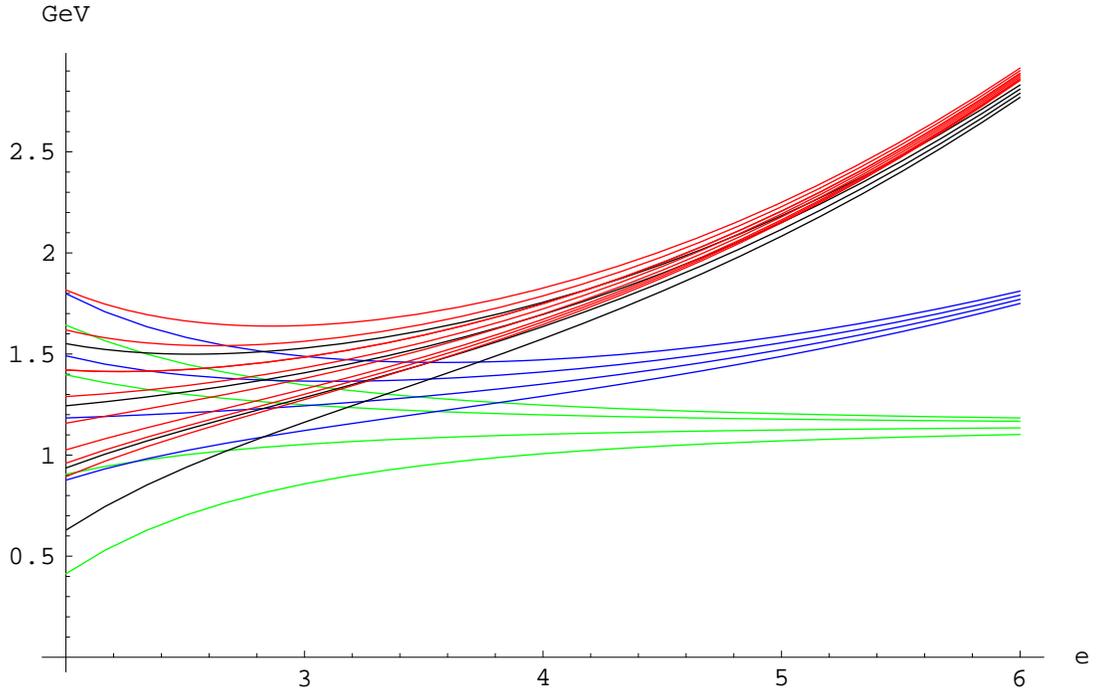}}
\caption{Fit (ii): Mass spectrums for {\bf 8} (green), {\bf 10} (blue), 
$\overline{\bf 10}$ (black) and ${\bf 27}_{3/2}$ (red).}
\label{fig6}
\end{figure}

\begin{figure}[h]
\centerline{\includegraphics{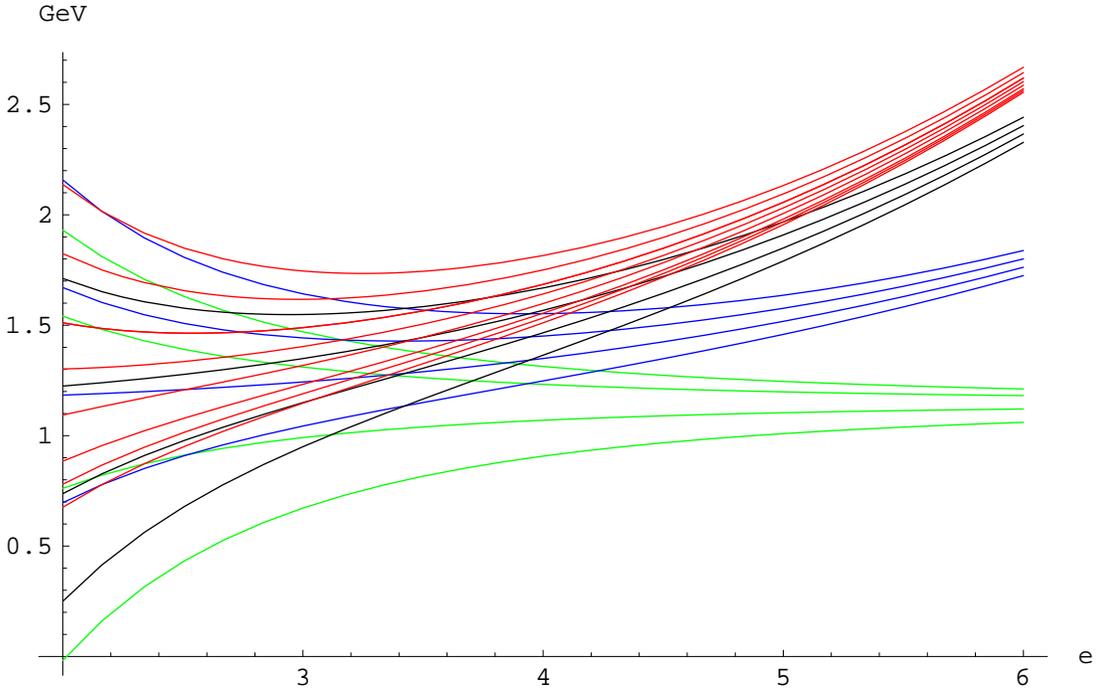}}
\caption{Fit (iii): Mass spectrums for {\bf 8} (green), {\bf 10} (blue), 
$\overline{\bf 10}$ (black) and ${\bf 27}_{3/2}$ (red).}
\label{fig7}
\end{figure}

\begin{figure}[h]
\centerline{\includegraphics{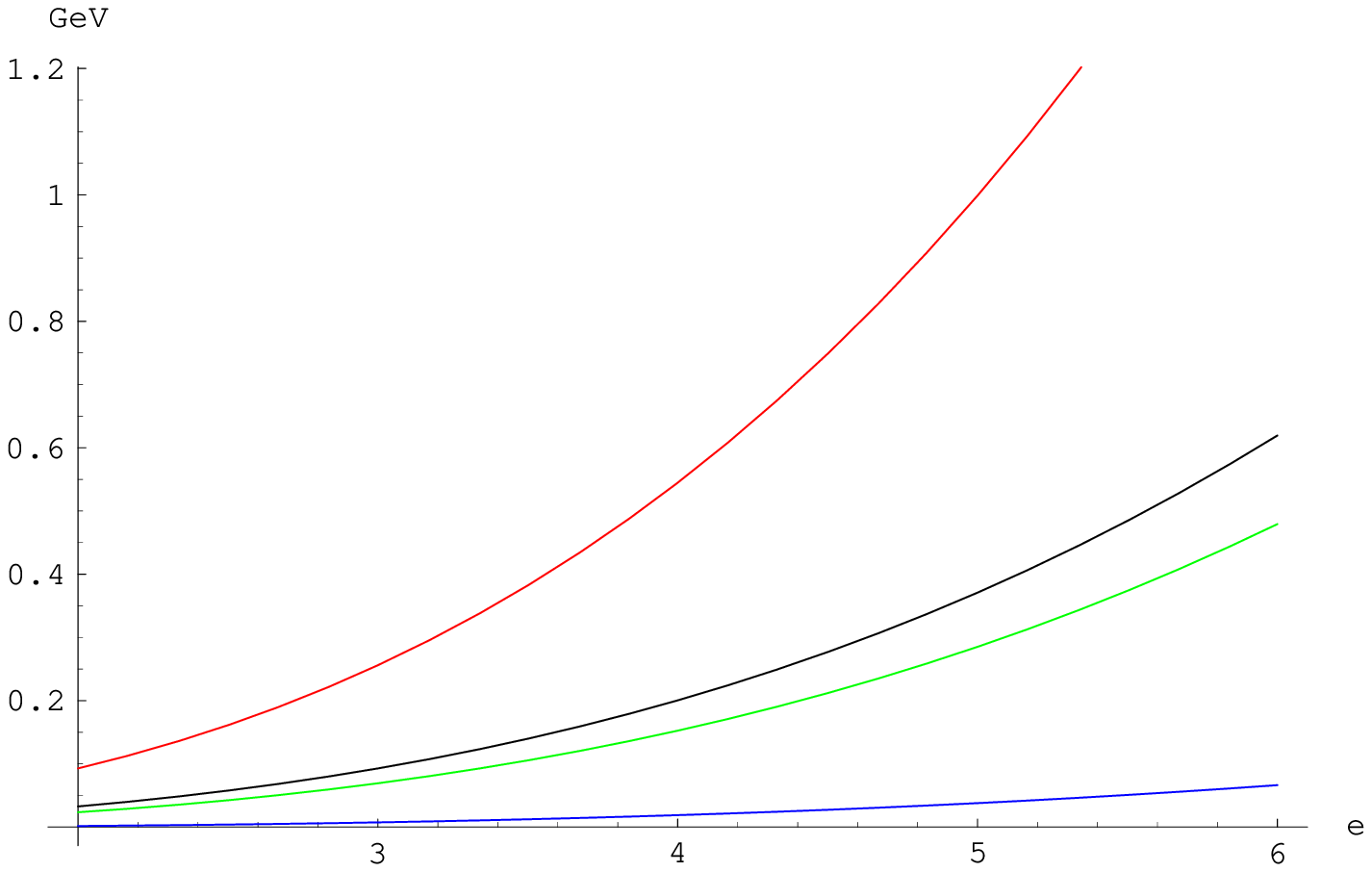}}
\caption{Fit (ii): Mean mass splittings $\Delta_3$ (blue), $\Delta_4$ (green), $\Delta_1$ (black) and $\Delta_2$ (red).}
\label{fig8}
\end{figure}

\begin{figure}[h]
\centerline{\includegraphics{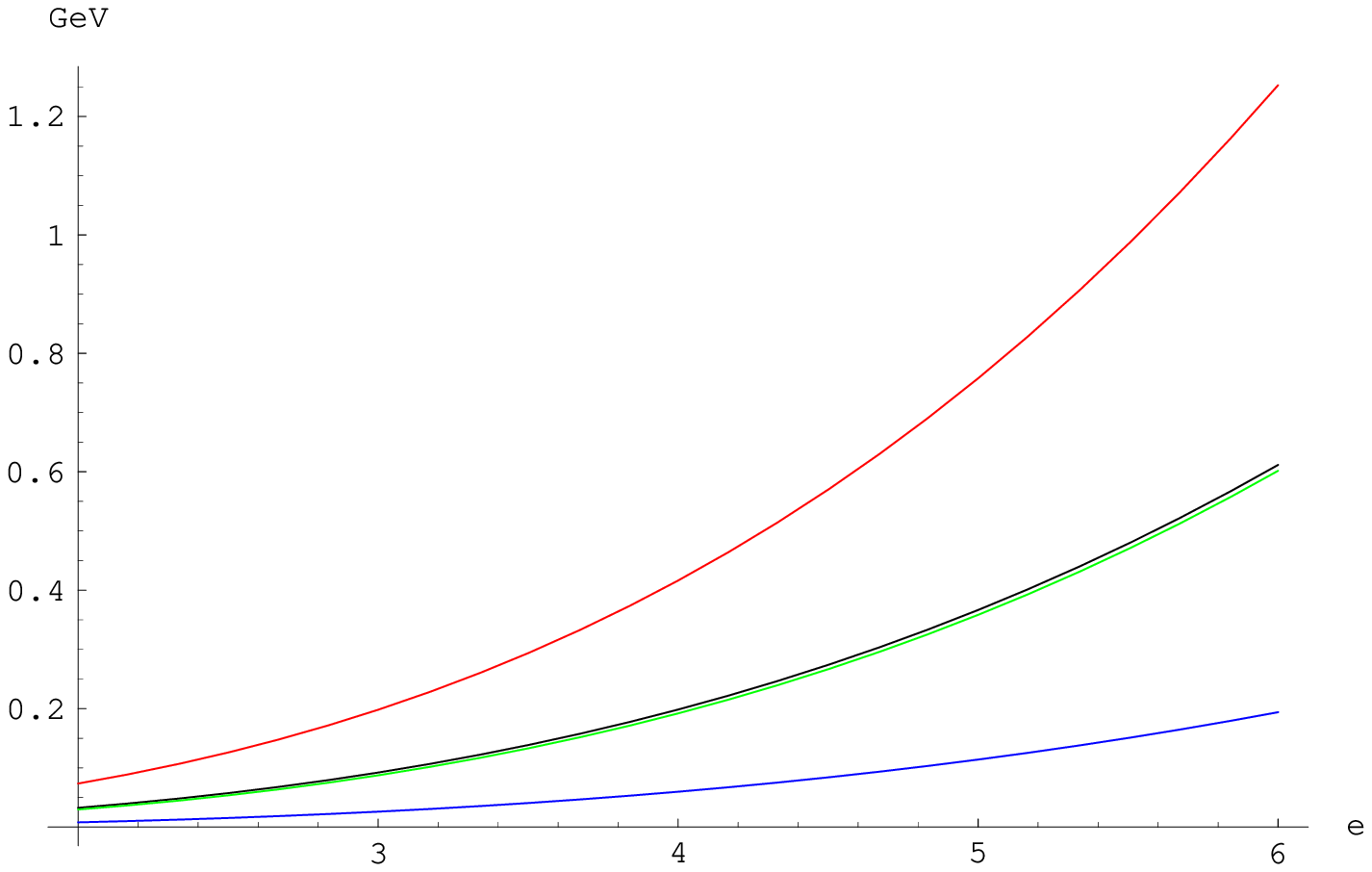}}
\caption{Fit (iii): Mean mass splittings $\Delta_3$ (blue), $\Delta_4$ (green), $\Delta_1$ (black) and $\Delta_2$ (red).}
\label{fig9}
\end{figure}

\begin{figure}[h]
\centerline{\includegraphics{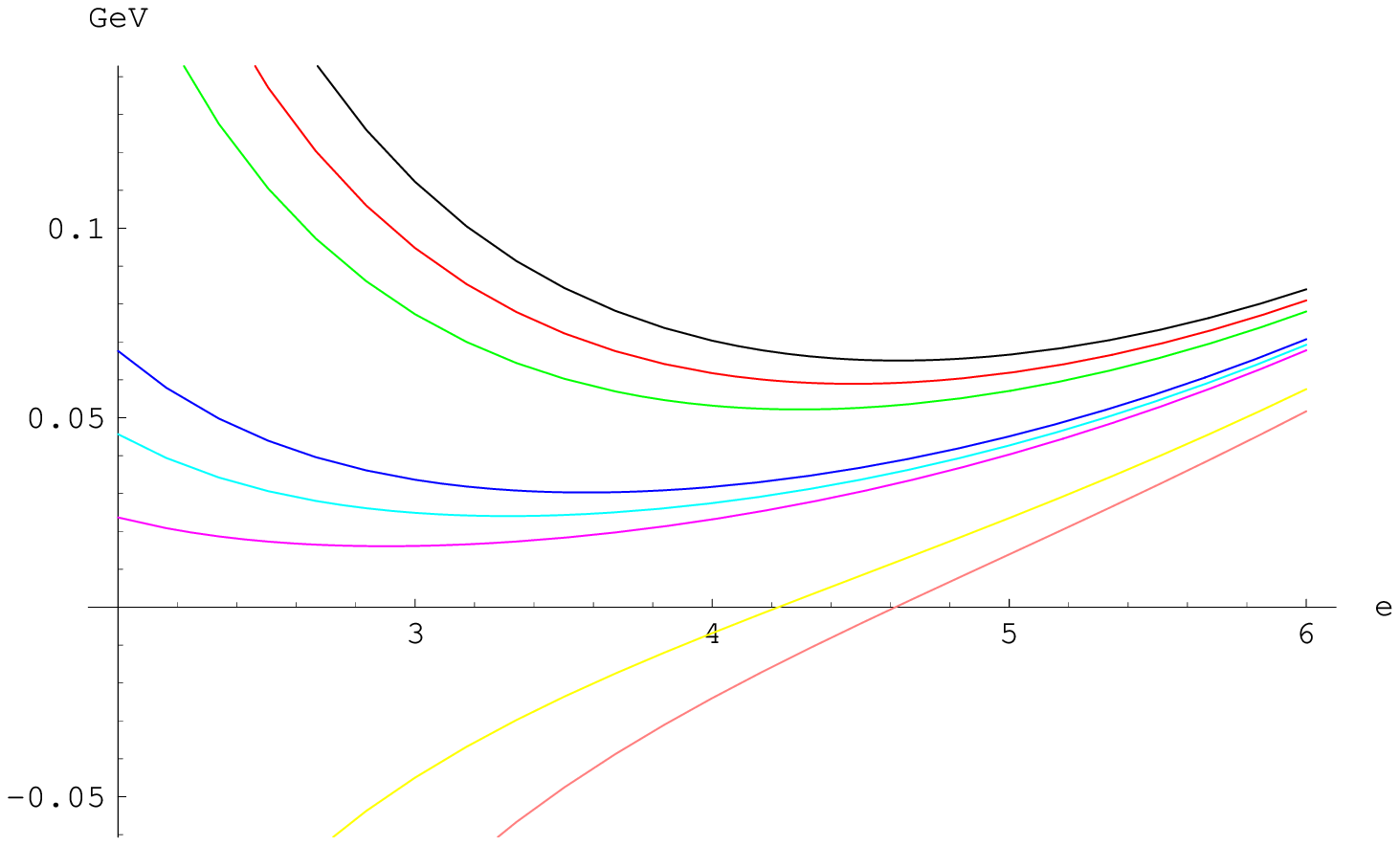}}
\caption{Fit (ii): Mass splittings $\delta_{4,7,2,5,8,6,3,1}$ - brown, yellow, pink, 
light-blue, dark-blue, green, red and black, respectively.}
\label{fig10}
\end{figure}

\begin{figure}[h]
\centerline{\includegraphics{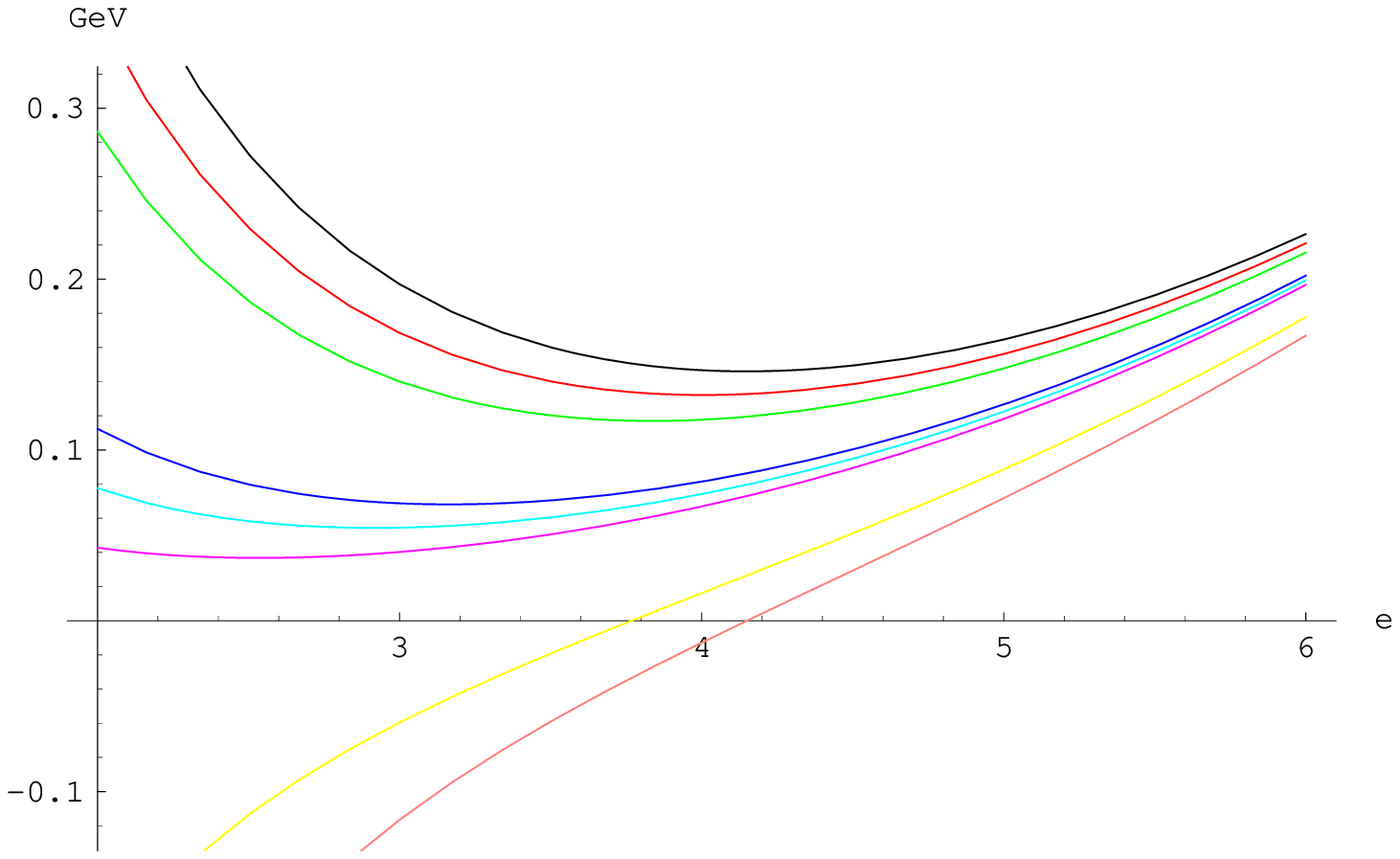}}
\caption{Fit (iii): Mass splittings $\delta_{4,7,2,5,8,6,3,1}$ - brown, yellow, pink, 
light-blue, dark-blue, green, red and black, respectively.}
\label{fig11}
\end{figure}

The influence of symmetry breaking effects, within 
minimal SU(3)$_f$ extended Skyrme model, on the prediction of 
the nucleon axial-vector current matrix element $g_A$,
the {\bf 8}, {\bf 10}, $\overline{\bf 10}$ and ${\bf 27}_{3/2}$ absolute mass spectrums and 
on the higher SU(3)$_f$ representation mass splittings
$\Delta_i,\, i=1,...,12$, for the multiplets 
${\bf 8},{\bf 10},\overline{\bf 10},{\bf 27},\bf{35}$, $\overline{\bf 35}$ and $\bf 64$ are important.
Their internal dynamics is, in the minimal approach with arctan ansatz 
for the profile function, described by the Eq. (\ref{86}).
Our Tables \ref{t:tab4} to \ref{t:tab9}, in comparison with the relevant 
numerics from \cite{wei1,dia1,WK,K,pra3,WM}, show implicitly that 
the inclusion of additional effects, like vector-mesons, 
the so-called static fluctuation and vibrations of Kaons \cite{wei} and other 
fine--tuning effects into the SB action \cite{WK,K,wei}
represents contributions of the order of a few percent and
does not change the conclusions dramatically. On the contrary, 
the main effect is due to the presence of $D_{88}$ term.
The importance of symmetry breaking effects has been  
demonstrated transparently in Figures \ref{fig2}-\ref{fig11}.
Our approach is similar to the one of \cite{WK,K}. The main difference is that our
action is simpler, i.e. it contains only symmetry breaking 
proportional to $\lambda_8$, and that we are using the 
arctan ansatz approximation for the profile function $F(r)$. 

For the axial-vector current matrix element $g_A$ with increasing symmetry breaking (\ref{87}), 
the two flavor result (\ref{103}) is slowly approached, Figs. \ref{fig2}-\ref{fig4}.
Using SU(3)$_f$ arctan ansatz for the profile function $F(r)$ and 
including next-to-leading terms like the Wess-Zumino term (\ref{52}) and
the SB term (\ref{53}), for $e=4$ we obtain $g_A = 0.97$ (Table \ref{t:tab4}).
This is about 23\% below the experimental value $g_A^{\rm exp} = 1.26$ and about 26\% below
our value of $g_A$ obtained for the pure 2-flavor case (\ref{103}), but within 1\% in agreement 
with value $g_A=0.98$ from Ref. \cite{wei}.
For $e=3.4$ equation (\ref{87}) gives $g_A=1.25$, 
in excellent agreement with experiment. However, it is understood 
that the explanation of the absolute mass spectrums with such low $e$-value  
is unreliable (see Figs. \ref{fig5}-\ref{fig7}).

Assuming equal spacing for antidecuplets, 
from the recent experimental data ($M^{\rm exp}_{\Theta^+}=1540$ MeV 
and $M^{\rm exp}_{\Xi^{--}_{3/2}}=1861$ MeV), 
in Ref. \cite{DT} we have found the following
masses of antidecuplets $M_{N^*}=1647$ MeV, 
$M_{\Sigma_{\overline{10}}} =1754$ MeV, the mean mass  
${\cal M}^{\overline{\bf 10}}=\frac{1}{10}\sum^{10}_{B=1}{M}^{\overline{\bf 10}}_B =1754$ MeV
and mass difference ${\Delta}_2$= 603 MeV. From Table \ref{t:tab8}, we also see that a
certain value of $e(=4.2)$ supports the case (ii), i.e. in good agreement with experiment. 
Taking $\Delta_2=$603 MeV together with ${\Delta}_1^{\rm exp}$, 
via Eq. (\ref{99}), we estimate ${\Delta}_3=30$ MeV, bonafide, as an ``experimental'' value.
It turns out that only $e\,\simeq\,3.2$, 
in the most realistic case (iii), could account for such small value of 
${\Delta}_3$. However $e=3.2$ gives to small values for ${\Delta}_1$ and ${\Delta}_2$.
Until now the only quantities which has required relatively small
value of Skyrme charge $e(\,\simeq\,3.4)$ in the 
minimal SU(3)$_f$ extended Skyrme model, was nucleon axial-vector coupling $g_A$ \cite{dppt}. 
Using 1754 MeV for the $\overline{\bf 10}$-plet mean mass and
predicted range for the mean mass splitting 
$30\,\leq \,{\Delta}_3\,\leq \,95$ MeV, we find the
range for the ${\bf 27}_{3/2}$-plet mean mass 
$1784\,\leq {\cal M}^{\bf 27}_{3/2}\,\leq 1849$ MeV, which is near the center of 
the ${\bf 27}_{3/2}$-plet mass spectrum displayed in Fig. 4 of Ref. \cite{WK}
(for A and B fits), and in Fig. 4 of Ref. \cite{Ell}.

\begin{figure}[h]
\centerline{\includegraphics{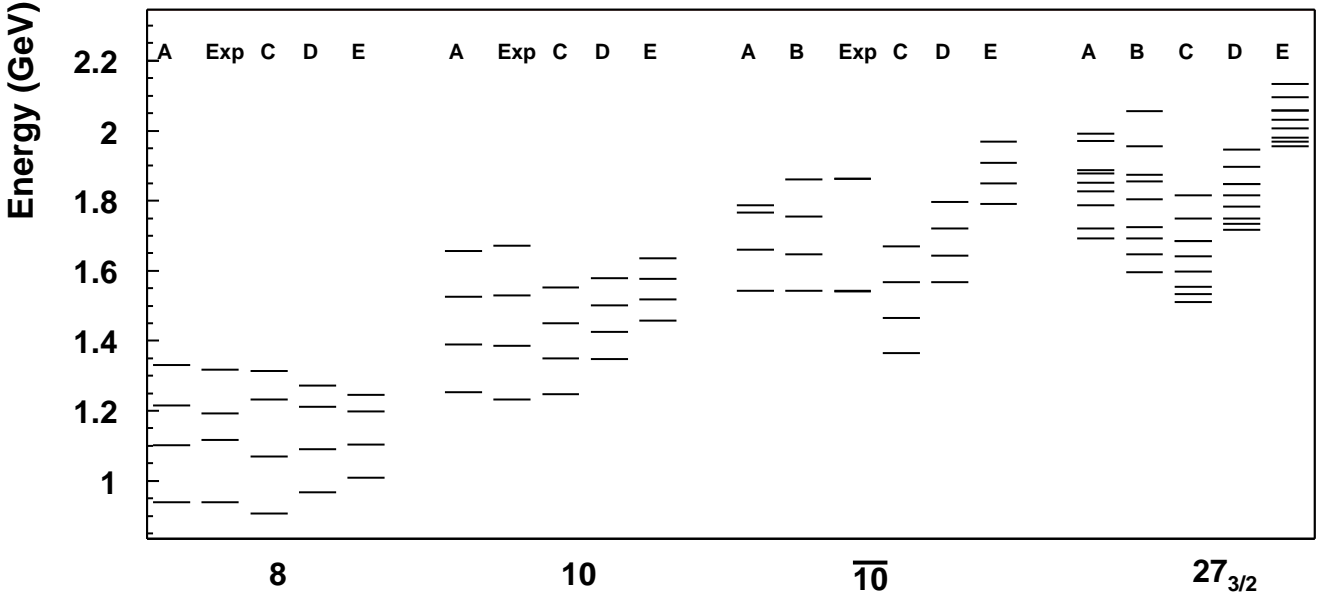}}
\caption{Graphical computation of the {\bf 8}, {\bf 10}, $\overline{\bf 10}$ and ${\bf 27}_{3/2}$ 
mass spectrums. Case A corresponds to the fit A from Fig. 4 in Ref. \cite{WK}.
Case B is Fig. 4 from \cite{Ell}. Cases C, D and E represents this paper for the fit (iii)
and for Skyrme charge $e=4.0;\,4.5;\,5.0$, respectively. For the $\overline{\bf 10}$ experimental masses we use
$M^{\rm exp}_{\Theta^+}=1540$ MeV and $M^{\rm exp}_{\Xi^{--}_{3/2}}=1861$ MeV.}
\label{fig12}
\end{figure}

Comparing the pure Skyrme model prediction for 
absolute masses of ${\bf 8},{\bf 10}$ and $\overline{\bf 10}$
in Ref. \cite{WK} (fits A and B in Table 2) with our results for the mass
spectrums (for $e=4.0;\,4.1$), presented in 
Tables \ref{t:tab6} and \ref{t:tab7}, we have found up to 8 \% discrapance. 
One of the reasons is that the fits A and B in Table 2 of Ref. \cite{K} were obtained for closed but
different $e$'s, i.e. for $e=3.96$ and $e=4.12$, respectively. 
Also, from Tables \ref{t:tab6} and \ref{t:tab7} one can see that for $e=4.2$, fit (iii), 
mass spectrum differs from the experiment $\leq 8$\% for 
$\Omega^-$, $\Theta^+$ and $\Xi_{3/2}^{--}$. 
Other estimated masses are $\leq\; 5$\% different from experiment.
Comparing our results for ${\bf 27}_{3/2}$ from Table \ref{t:tab7}, with the Skyrme model
prediction of Ref. \cite{WK} (fits A and B in Figure 4) 
shows that our case (iii) with $4.3\,\leq\,e\,\leq \,4.7$ supports the fit B, and
for $4.4\,\leq\,e\,\leq \,4.6$ agrees nicely with the fit A. Both fits A and B from \cite{WK}
lies between $4.0\,\leq\,e\,\leq \,4.6$ for case (ii).
Case (iii) with $4.2\,\leq\,e\,\leq \,4.7$ 
support also the results presented in Table 1. of Ref. \cite{WM}.
For the narrower $e$-range, $4.4 \leq e \leq 4.7$, 
the prediction for the $\overline{\bf 10}$ masses would
lie inside the following range of values $1526\leq \,M_{\Theta^+}\,\leq 1654$ MeV, 
$1607\, \leq M_{\rm N^*}\, \leq 1723$ MeV, $1687\, \leq M_{\Sigma_{\overline{10}}}\leq 1792$ MeV 
and $1768\,\leq \,M_{\Xi^{--}_{3/2}}\,\leq 1861$ MeV, respectively. 
From Tables \ref{t:tab6} and  \ref{t:tab7} we conclude that in the minimal approach the best fit for 
{\bf 27}$_{3/2}$-plet mass spectrum,
as a function of $e$ and for $f_K\not=f_{\pi}$, would lie between $e\simeq 4.2$ and $e\simeq 4.7$,
just like for the octet, decuplet and anti-decuplet mass spectrums \cite{DT}, and  
agree reasonably well with both baryon spectrums from Table 4.1 in Ref. \cite{wei} and from
Table 2 (fits A and B) of Ref. \cite{K}, respectively. 
In Table \ref{t:tab7} masses of $\Lambda^*$ and $\Xi^*_{\frac{3}{2}}$ are equal due to the absence
of anomalous moments of inertia \cite{MP,HB} in the model used in this paper.
Note, however, that anomalous moments of inertia 
contributions are parametrized in \cite{WM,Ell} to be at best 
$\sim$ 1 \% for the $\Xi^*_{\frac{3}{2}}$ mass, for example. 
Clearly, this way, they represent just the fine--tuning type of effects.
For a few fixed values of $e$ the mass spectrums of ${\bf 8},{\bf 10}$, $\overline{\bf 10}$ 
and ${\bf 27}_{3/2}$-plets are given in Figure \ref{fig12}.

The higher SU(3)$_f$ representation mass splittings $\Delta_i,\, i=3,...,12$, for the multiplets 
${\bf 8},{\bf 10},\overline{\bf 10},{\bf 27},\bf{35},\overline{\bf 35}$ and $\bf 64$ expressed in terms 
of decuplet--octet and antidecuplet--octet mass splittings $\Delta_1$ and $\Delta_2$ 
are given in (\ref{99}) and (\ref{101}). 
With the help of Table \ref{t:tab8}, from (\ref{99}) we obtain predictions for 
mass splittings $\Delta_i,\, i=5,...,12$,
as functions of different dynamical assumptions (\ref{83}) and the Skyrme charge $e$.
For example for case (iii), and for the range of $4.4\leq \,e\,\leq 4.7$
which fits well the antidecuplet masses,
we predict the following range for the mass splittings $536\,\leq \Delta_2\,\leq 641$ MeV
and $250\,\leq \Delta_4\,\leq 301$ MeV, respectively. 
It is important to stress that, since the mass splittings (\ref{99}-\ref{102}) 
depend on the inverse moments of inertia (\ref{88}) and (\ref{89}) only, 
i.e. there are no additional inputs of the same or higher power in $e$, and
consequently they scale as $\Delta^{\cal R'}_{\cal R}\sim e^3$,
the model predicting power for them is the most sensitive. This is illustrated 
in Figure \ref{fig12}, by the difference between cases C$(e=4)$, D$(e=4.5)$ and E$(e=5)$ 
for the ${\overline{\bf 10}}$ and ${\bf 27}_{3/2}$ spectrums, where
spectral lines follows notations from Tables \ref{t:tab6} and \ref{t:tab7}:
$({\bf 8})=({\rm N},\Lambda,\Sigma,\Xi)$, $({\bf 10})=(\Delta, \Sigma^*,\Xi^*,\Omega)$, 
$(\overline{\bf 10})=(\Theta^+,{\rm N}^*,\Sigma_{\overline{10}},\Xi^{--}_{3/2})$ 
and $({\bf 27}_{3/2})=(\Theta_1,{\rm N}^*_{\frac{3}{2}},\Sigma_2,{\rm N}^*_{\frac{1}{2}},\Sigma_1,
\Lambda^*,\Xi^*_{\frac{3}{2}},\Xi^*_{\frac{1}{2}},\Omega_1)$, in accord with
notation in Fig. 4 from \cite{Ell}.
Note that in Figure \ref{fig12} column A, for ${\bf 27}_{3/2}$-plet, 
(extracted from Fig. 4 column A of \cite{WK}) 
the state $\Xi^*_{\frac{3}{2}}\equiv (-1,3/2)$ lies below state $\Sigma_1\equiv (0,1)$,
due to the absence of the configuration mixing in the evaluation of the ${\bf 27}_{3/2}$-plet
spectrum in this paper and in Ref. \cite{Ell}.

Although we are using simple version of the total action (\ref{49}),
our results for the nucleon axial-vector coupling, moments of inertia, mass spectrum and mass differences
given in Tables \ref{t:tab4}--\ref{t:tab8},
do agree well with the other Skyrme model based estimates \cite{MP,dia1,wei1,WK,K,pra3,Itz,Ell,wei}. 
Careful inspection of the results for the ${\bf 27}_{3/2}$-plet mass spectrum from Fig. 4
of Ref. \cite{WK} also shows approximative agreement with our results, $\delta_1,...,\delta_8$,
for $4.2 \,\leq \,e\,\leq 4.7$ fit (iii), presented in 
Table \ref{t:tab9} and in Figs. \ref{fig10} and \ref{fig11}.
These numbers are in good agreement with the results obtained recently \cite{WK,K,WM,Ell,HW}.
On top of the importance of the $e$-dependence it turns out that
the dependence on the difference between $f_{\pi}$ and $f_K$ 
is crucial for the correct description of the small mass splittings in (\ref{99}). 
For the small mass splittings, like $\Delta_3$, the contribution of 
the term proportional to $(f^2_K-f^2_{\pi})$ in
the SU(3)$_f$ symmetry breaking term ${\cal L}_{\rm SB}$ plays a major role.
It is clear from Figs. \ref{fig2}-\ref{fig11} that SB effects are sizeable 
and change the relevant quantities. The exception is the $g_A$ which change modestly.

The ${\bf 27}_{3/2}$--$\overline{\bf 10}$ mass splittings, given in Tables 
\ref{t:tab8} and \ref{t:tab9}, are quantities whose 
measured values, together with measurements of the decay modes branching ratios and relevant widths, 
would determine the spins 3/2 or 1/2, of observed objects like $\Xi^{--}_{3/2}$, 
placing it into the right SU(3)$_f$ representation 
$\overline{\bf 10}$, ${\bf 27}_{3/2}$ or ${\bf 27}_{1/2}$.
We do expect that experimental analysis, considering other members of the ${\bf 27}_{3/2}$ and 
$\overline{\bf 10}$-plets, should be performed in the near future. 
Since the mass splittings $\Delta_3$ represent 
the smallest splittings among all of the splittings between 
the SU(3)$_f$ multiplets ${\bf 8}$, ${\bf 10}$, $\overline{\bf 10}$, ${\bf 27}$, ${\bf 35}$, 
$\overline{\bf 35}$ and ${\bf 64}$
we would urge our colleagues to continue present penta-quark spectral and decay modes experimental analysis 
and find the penta-quark members of the ${\bf 27}_{3/2}$-plet which would mix with or 
lie just above the penta-quark family of the $\overline{\bf 10}$-plet.
All mentioned experiments would finally show which model, quark or soliton in general,
is better describing penta-quark, septu-quark, etc. states. 
However, one might speculate that the correct description of those states lie somewhere in between.

We hope that the present calculation, taken together with 
the analogous calculation in \cite{wei,WK,K,pra3,WM,Ell,DT} will
contribute to understanding of the overall picture of 
the baryonic mass spectrum and mass splittings in the Skyrme model,
as well as for further computations of other non perturbative, dimension-6 operator matrix elements between
different baryon states \cite{dppt,dppt1,tra,prat}.


\vspace{0.1cm}
We would like to thank T. Anti\v ci\' c and K. Kadija for helpful discussions.
One of us (JT) would like to thank V.B. Kopeliovich, A.V. Manohar, M. Praszalowicz 
and J. Wess for stimulating discussions and Theoretische Physik, Universit\" at
M\" unchen and Theory Division CERN, where part of this work was done, for hospitality.
This work was supported by the Ministry of Science and Technology of the Republic of Croatia under Contract 0098002.






\end{document}